\documentclass[aps,prd,preprint,superscriptaddress,tightenlines,nofootinbib,floatfix]{revtex4}
\usepackage{epsfig}
\usepackage{graphicx}
\usepackage{dcolumn}% Align table columns on decimal point
\usepackage{bm}% bold math

\begin{document}
\def\etal{{\it et al.}}
\newcommand{\jpsi}{J/ \psi}
\def\upsln{\Upsilon (1S)}
\def\jpsix{\jpsi+X}
\def\jpsiee{\jpsi\to e^+e^-}
\def\jpsimm{\jpsi\to \mu^+\mu^-}
\def\jpsill{\jpsi\to l^+l^-}
\def\upstojpsix{ \Upsilon (1S)\to \jpsi+X}
\def\upstochicx{ \Upsilon (1S)\to \chi_{cJ}+X}
\def\upstochic0{ \Upsilon (1S)\to \chi_{c0}+X}
\def\upstochic1{ \Upsilon (1S)\to \chi_{c1}+X}
\def\upstochic2{ \Upsilon (1S)\to \chi_{c2}+X}
\def\upstojpsiccg{ \Upsilon (1S)\to \jpsi c\bar{c}g+X}
\def\upstoqq{ \Upsilon (1S)\to q\bar{q}}
\def\upstomm{ \Upsilon (1S)\to\mu^+\mu^-}
\def\eetoqq{ \Upsilon (1S)\to q\bar{q}+X}
\def\eetomm{e^+e^-\to \mu^+\mu^-}
f
\def\eetojpsix{e^+ e^-\to\jpsi+X}
\def\ups4s{ \Upsilon (4S)}
\def\upsl4{ \Upsilon (4S)}
\def\psitwos{ \psi (2S) }
\def\ifb{\rm fb^{-1}}

\preprint{CLNS 04/1883}
\preprint{CLEO 04-08}

%\preprint{\tighten\vbox{\hbox{\hfil CLNS {04/1883}}
%                        \hbox{\hfil CLEO {04-08}}
%}}

%\draft % needed for PACS to appear
%\tighten

\title{New Measurements of $\upsln$ Decays to Charmonium Final States}

\author{R.~A.~Briere}
\author{G.~P.~Chen}
\author{T.~Ferguson}
\author{G.~Tatishvili}
\author{H.~Vogel}
\author{M.~E.~Watkins}
\affiliation{Carnegie Mellon University, Pittsburgh, Pennsylvania 15213}
% ...

%\author{(CLEO Collaboration)} %FOR PRD
\collaboration{CLEO Collaboration} %FOR PRL,CLNS
\noaffiliation

% You will want to hard code the date once you are ready to submit your paper!
\date{Oct. 5, 2004}

\begin{abstract} 
Using data collected by the CLEO III detector at CESR, we report on measurements of
$\upsln$ decays to charmonium final states. The data sample used for this analysis consists of 
$21.2\times 10^6$
$\upsln$ decays, representing about 35 times more data than previous
CLEO $\upsln$ data samples. We present substantially improved measurements of
the branching fraction ${\cal{B}}(\upstojpsix)$ using $\jpsimm$ and $\jpsiee$
decays. The branching fractions for these two modes are averaged,
thereby obtaining: $
{\cal{B}}(\upstojpsix) = (6.4\pm0.4$(stat)$\pm$0.6(syst))$\times 10^{-4}$. 
A greatly improved measurement of the $\jpsi$ momentum distribution 
is presented and indicates a spectrum which is much softer than predicted 
by the color-octet model and somewhat softer than the color-singlet model.
First measurements of the
$\jpsi$ polarization and production angle are also presented. In addition, we report 
on the first observation of $\upsln\to\psitwos+X$ and evidence for $\upstochicx$. 
Their branching fractions are measured relative to ${\cal{B}}(\upstojpsix)$
and are found to be: 
${{\cal{B}}(\upsln\to\psitwos+X)\over {\cal{B}}(\upstojpsix )} = 0.41\pm0.11$(stat)$\pm$0.08(syst),
${{\cal{B}}(\upsln\to\chi_{c1}+X)\over {\cal{B}}(\upstojpsix)}$ = 0.35$\pm$0.08(stat)$\pm$0.06(syst), 
${{\cal{B}}(\upsln\to\chi_{c2}+X)\over {\cal{B}}(\upstojpsix)}$ = 0.52$\pm$0.12(stat)$\pm$0.09(syst), and
${{\cal{B}}(\upsln\to\chi_{c0}+X)\over {\cal{B}}(\upstojpsix)} < 7.4$ at 90\% confidence
level. 
The resulting feed-down contributions to $\jpsi$ are (24$\pm$6(stat)$\pm$5(syst))\% for
$\psitwos$, (11$\pm$3(stat)$\pm$2(syst))\% for $\chi_{c1}$, (10$\pm$2(stat)$\pm$2(syst))\% 
for $\chi_{c2}$, and $<8.2\%$ at 90\% confidence level for $\chi_{c0}$. These measurements
(apart from $\chi_{c0}$) are about a factor of two larger than expected based on the 
color-octet model. 
\end{abstract}

\maketitle
%\tighten

\setcounter{footnote}{0}

\newpage
%\section{Introduction}\label{sec:Introduction}

\section{\boldmath Introduction}\label{sec:Introduction}

	Charmonium has played a crucial role in the recent history of particle physics.
It has been nearly 30 years since its discovery in both $e^+e^-$
interactions~\cite{jpsi_discovery1} and in collisions of protons on a beryllium target
~\cite{jpsi_discovery2}. Over the last two decades, the charmonium and
bottomonium systems have served as a laboratory for testing QCD. In the weak sector, charmonium 
also serves as a critical tool in extracting CKM~\cite{ckm} phases in $B$-meson decays. 
However, even after 
30 years of studying $c\bar{c}$ systems, we still lack a complete understanding of their
production mechanisms in glue-rich environments. 

	About a decade ago, the CDF experiment reported production rates of charmonium
in proton-antiproton collisions which exceeded the existing theoretical calculations by
a factor of about 10 for $\jpsi$ and about a factor of 50 for $\psitwos$~\cite{cdf1}.
An explanation of this excess was given by the so-called {\it color-octet} 
mechanism~\cite{color_octet_mech}, whereby a gluon fragments into a color-octet 
$^3S_1$ $c\overline{c}$ pair, which then
evolves non-perturbatively into a color-singlet by emission of a soft gluon. 
The size of this non-perturbative matrix element is not predicted and was 
determined by a fit to the CDF data. Because of the
glue-rich environment and the $\alpha_s^2$ suppression of the color-singlet process
relative to the color-octet process, it was argued~\cite{color_octet_mech} that
the latter contribution is likely to be important. While this model can explain
the rate and momentum spectrum of $\jpsi$ and $\psitwos$ production at the Tevatron
it appears that it does not properly describe recent $\jpsi$ polarization data from 
CDF~\cite{cdf2}. Furthermore, when the same matrix elements determined at CDF are
applied to photoproduction of $\jpsi$ at HERA, the color-octet 
contribution is about a factor of ten too large~\cite{h1}.

	Over the last several years, the role of the color-octet mechanism in $\jpsi$
production in $e^+ e^-$ collisions has been under theoretical study~\cite{ee_jpsi}. 
The dynamics of the color-octet processes are expected to give rise to significant 
differences in the $\jpsi$ momentum spectrum and production angle as compared to 
color-singlet production. Recently, both BaBaR~\cite{babar_ee_jpsi}
and Belle~\cite{belle_ee_jpsi} have reported measurements of the cross section and
the momentum spectra of $\jpsi$'s in $e^+e^-$ collisions on the $\ups4s$ ($B$ decays
excluded) or just 
below the $\upsl4$ ({\it i.e.,} in the continuum). BaBaR measures 
$\sigma(e^+e^-\to \jpsi+X)=(2.52\pm0.21\pm0.21$) pb, 
whereas Belle measures a number which is 40\% lower, (1.47$\pm$0.10$\pm$0.13) pb (about
3 standard deviations below than the BaBaR result). The two experiments both
observe similar shapes for the $\jpsi$ momentum spectrum, which are 
softer than the predictions of the color-octet model~\cite{ee_jpsi} which predict
a peaking of the $\jpsi$ momentum spectrum near the kinematic endpoint. 
However, recent theoretical studies of the color-octet subprocesses,
$e^+e^-\to\jpsi+g$~\cite{fleming1} and $e^+e^-\to\jpsi+gg$~\cite{lin-zhu},
show that the perturbative expansion breaks down near the kinematic
endpoint, and the authors appeal to soft-collinear effective theory (SCET)
to systematically include the non-perturbative effects. In Ref.~\cite{fleming1}, 
it is shown that by using SCET
the color-octet model predictions can be sufficiently softened 
and reasonably good agreement with $\eetojpsix$ data can be achieved, 
although the calculation is not completely predictive
because it uses a shape function which is fit to the $\eetojpsix$ 
data~\cite{belle_ee_jpsi,babar_ee_jpsi}. Belle also
reports on production of $\psitwos$ in $e^+e^-$ collisions, with
a measured ratio  
$\sigma(e^+e^-\to \psitwos+X) / \sigma(e^+e^-\to \jpsi_{{\rm direct}}+X) = 0.93\pm0.17^{+0.13}_{-0.15}$.
~\cite{belle_ee_jpsi}.
That is, the production rates for $\jpsi$ and $\psitwos$ in $e^+e^-$ collisions are
approximately equal. The color-singlet mechanism can yield the $\psitwos$ final state, 
but the expected ratio is $\leq$10\%~\cite{li-xie-wang}.
Belle has extended their inclusive $\jpsi$ analysis to search for associated charmed particles,
and they find: $\sigma(e^+e^-\to \jpsi c\overline{c}) / \sigma(e^+e^-\to\jpsi+X) 
= 0.59^{+0.15}_{-0.13}\pm0.12$~\cite{belle_ee_jpsi2}.  The
color-octet contribution is expected to be at the level of 
1\%~\cite{cheung_1s} of the inclusive rate.
The disagreement here indicates that the production mechanisms of charmonium
are not well understood, and more theoretical and experimental input is required.

	Several theoretical papers~\cite{cheung_1s,napsuciale} have suggested that 
the study of $\jpsi$ production in $\upsln$ decays could provide an alternate probe 
of the charmonium system in that the $\upsln$ decay provides a
glue-rich environment in which $\jpsi$ mesons can be produced abundantly through
the color-octet mechanism. The kinematics of such $\jpsi$'s are expected to exhibit
signatures distinct from other production mechanisms, such as a peak in
the $\jpsi$ momentum spectrum near the kinematic endpoint. The predicted 
branching fraction from color-octet processes is 
${\cal{B}}(\upstojpsix)=6.2\times 10^{-4}$~\cite{napsuciale},
with approximately 10\% feed-down expected from $\psitwos$ and another 
10\% from $\chi_{cJ}$~\cite{trottier} (summed over all $J$). 
Color-singlet processes, 
such as $\upsln\to\jpsi+gg$ start at $\alpha_s^6$, and are therefore suppressed
relative to color-octet processes which enter at $\alpha_s^4$.
However, computations of the color-singlet process $\upstojpsiccg$ ~\cite{li-xie-wang}
indicate a sizeable branching fraction of $5.9\times 10^{-4}$, with about 10\% 
coming from $\psitwos$ feed-down. The enhancement here arises because 
the non-perturbative color-singlet matrix element for $c\bar{c}\to\jpsi$
may be 210-360 times larger than the corresponding color-octet matrix element, 
which is enough to compensate
for the perturbative suppression. Moreover, unlike the color-octet processes,
this process inherently results in a soft $\jpsi$ momentum spectrum because
of the two additional charm quarks in the final state. As a result, the
$\jpsi$ momentum cannot exceed about 3.3 GeV/$c$ in this process.
Therefore, while the color-octet and color-singlet processes give similar
predictions for total rate, their momentum distributions are significantly
different. Figure~\ref{jpsi_feyn2} shows Feynman diagrams for (a) two of the
more important color-octet processes and (b) the $\upstojpsiccg$ color-singlet 
diagram. It should be noted that color-singlet production would also be 
signalled by the presence
of additional charmed particles (open charm) in association with the $\jpsi$.
To capitalize on the small yield of $\jpsi$'s in $\upsln$ decay, many $D$ decay
channels, both inclusive and exclusive will need to be explored. We therefore
relegate the search for open charm in association with $\jpsi$ in $\upsln$ decay
to a future report. 

\begin{figure}[btp]
\centerline{
\includegraphics[width=5.0in]{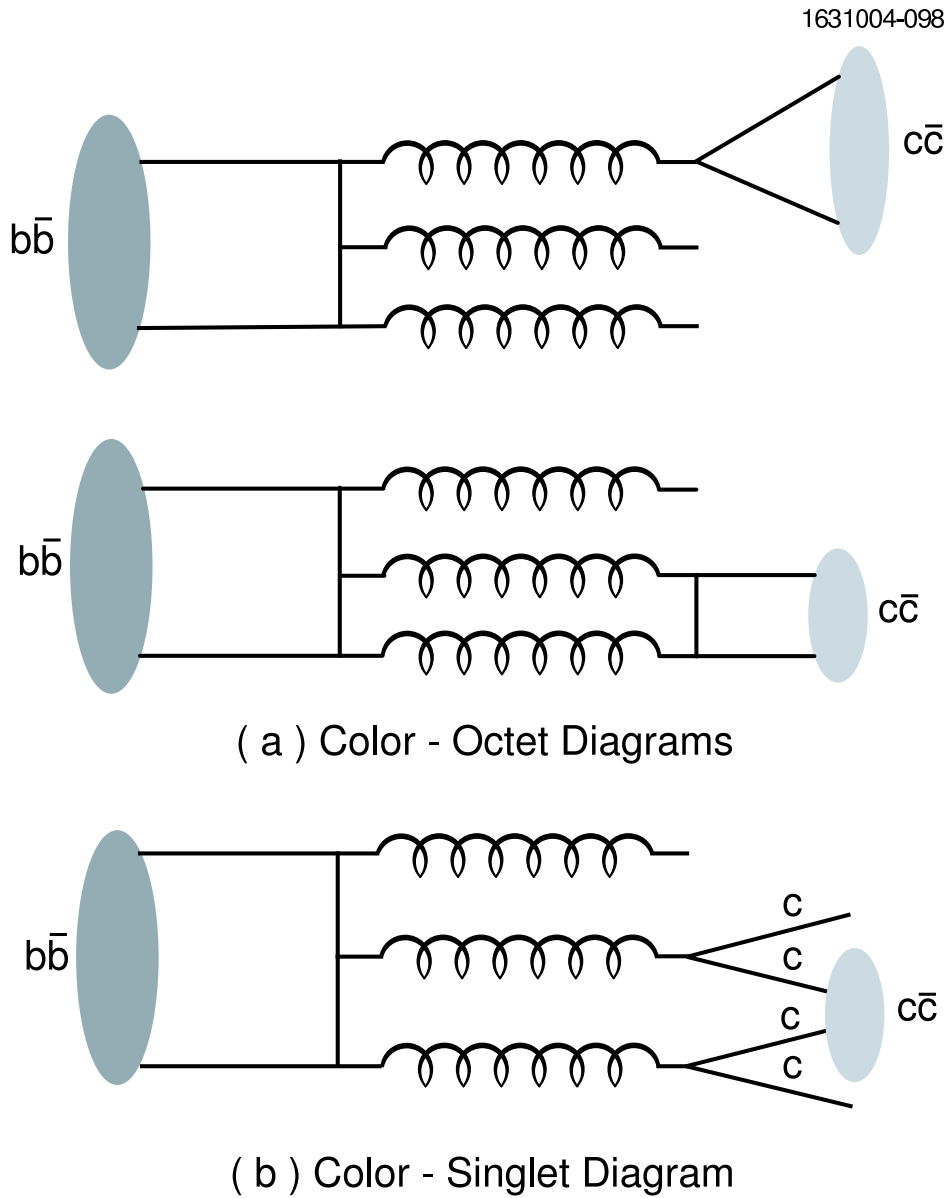}}
\caption{\label{jpsi_feyn2} 
Feynman diagrams for production of charmonium in $\upsln$ decays from
(a) color-octet processes and (b) color-singlet $\upstojpsiccg$. For the
color-octet processes, the $\jpsi$ is produced in a color-octet and becomes
a color-singlet through emission of a soft gluon.}
\end{figure}

	The process $\upstojpsix$ has been previously observed by 
CLEO~\cite{cleo-upstojpsix}, where the branching fraction was measured
to be (1.1$\pm$0.4$\pm$0.2)$\times 10^{-3}$ based on $\approx$20 observed
events. CLEO also reported a soft momentum spectrum for the $\jpsi$,
albeit with limited statistical precision.
The ARGUS Collaboration reported an upper limit of 
0.68$\times 10^{-3}$~\cite{argus-upstojpsix} at 90\% confidence level. 

	The CLEO Collaboration has collected large data samples on the $\Upsilon$(n$S$) 
resonances and currently has the world's 
largest samples of $\upsln$, $\Upsilon$(2$S$) and $\Upsilon$(3$S$) decays. 
Consequently, CLEO is in a unique position to help clarify the roles of 
color-singlet and color-octet models in $\jpsi$ production. 

	In this paper, we present vastly improved measurements of the rate,
momentum spectrum, and angular distributions in $\upstojpsix$ decays. 
We also present first observations 
of the decays $\upsln\to\psitwos+X$ and evidence for $\upsln\to\chi_{c1,2}+X$. 
The paper is organized as follows. In Section~\ref{sec:evsel} we discuss
the data samples used, the $\jpsi$ backgrounds, event selection and $\jpsi$
reconstruction. Section~\ref{sec:upstojpsix} details the measurement of the
$\upstojpsix$ branching fraction and momentum spectrum in $\upsln$ decays. This section also
includes a measurement of the cross section $\sigma(\eetojpsix)$ using data on and 
below the $\upsl4$ resonance, which is used 
to estimate and subtract the continuum contribution at the $\upsln$. The section
concludes with an examination of some event-level distributions. Section~\ref{sec:psitwos}
presents the measurement of the $\upsln\to\psitwos+X$ branching fraction.
The report then discusses in Section~\ref{sec:upstochicx}
the measurement ${\cal{B}}(\upstochicx)$. For each of these analyses, we present
a cross-check by measuring the corresponding branching fraction in $B$-meson decay.
Lastly, we discuss in Section~\ref{sec:sys} the systematic uncertainties in each of 
these analyses, and the paper is concluded in Section~\ref{sec:summary}.

\section{\boldmath Data Samples, Backgrounds, Event Selection, and $\jpsi$ Reconstruction}
\label{sec:evsel}

	The analysis presented here uses data collected using the CLEO III 
detector~\cite{cleoiii}.
The primary data sample includes 1.2 $\ifb$ of data collected on the $\upsln$, and
amounts to (21.2$\pm$0.2)$\times 10^6$ $\upsln$ decays. For background determinations
and systematic checks, we also utilize
5.0 $\ifb$ of data on the $\upsl4$ resonance (10.4 million $B$-meson decays)
and 2.3 $\ifb$ just below ($\approx$10.56 GeV) the $\upsl4$ resonance. We also
use the on-$\upsl4$ data for cross-checks on charmonium yields in $B$-meson decays.

%A summary of these data sets is shown in Table~\ref{datasets}.
%
%\begin{table*}[hbt] 
%\begin{center}  
%\caption{Summary of data sets used for this analysis.~\label{datasets}} 
%\begin{tabular}{|c|c|c|c|}\hline 
%Type & Beam  Energy (GeV) &  Luminosity ($\ifb$) & \# Decays (Type) \\
%\hline\hlineE eV
%$\upsln$ &  9.46 & 1.2 & (21.2$\pm$0.2)$\times 10^6$ ($\Upsilon$(1S))    \\
%\hline
%Below $\upsln$ & 9.44 &  0.2 & - \\
%\hline
%On $\upsl4$    & 10.58 & 5.0  & ($10.4\pm 0.5)\times 10^6$ ($B$ mesons) \\
%\hline
%Below $\upsl4$   & 10.56 & 2.3  & - \\
%\hlin$e 
%\end{tabular} 
%\end{center} 
%\end{table*} 

	The backgrounds to $\jpsi\to l^+l^-$ on the $\upsln$ are: 
(a) radiative Bhabha events, (b) $\gamma\gamma$ fusion producing $\chi_{cJ}$ which 
subsequently produces $\jpsi\gamma$, (c) radiative return processes such as
$e^+e^-\to\jpsi\gamma$ or $e^+e^-\to\psitwos\gamma$, and (d) continuum production
($\eetojpsix$). 
Various event selection requirements are targeted at reducing or
eliminating these backgrounds. Radiative Bhabha events produce background
in the $\jpsi$ mass region when one of the hard leptons is combined with
a soft lepton from the converted photon. Such events are suppressed 
by requiring that the invariant mass of either electron from the
$\jpsiee$ candidate with any other electron in the event have $M_{ee}>100$ MeV/$c^2$. 
Events produced through $\gamma\gamma\to\chi_{cJ}$ fusion typically only leave 
two charged tracks in 
the CLEO III detector, and these events are therefore easily 
rejected by a requirement of at least 3 charged tracks. The radiative
return backgrounds are suppressed through event selection criteria 
which take advantage of the special kinematics of these processes, namely 
a low particle multiplicity coupled with either the detection of
a high energy photon ($\approx$4 GeV) or large missing event momentum.
Events are required to have their missing event momentum magnitude, $P_{\rm ev}<3.75$ GeV/$c$, 
or, if the number of charged tracks, $N_{\rm trk}\le 4$, we require 
$P_{\rm ev}<2.0$ GeV/$c$. When
the high energy photon is detected (or an $e^+e^-$ pair with
invariant mass less than 100 MeV/$c^2$), the event is vetoed if
$N_{\rm trk}\le 4$ and the (converted) photon has energy greater than 3.75 GeV.
The remaining background from these three sources to the $\upstojpsix$ 
signal is negligible. However, because of the small signal in 
$\upsln\to\psitwos+X$, the remaining background cannot be neglected.
This background is determined using the {\sc evtgen} MC followed
by a GEANT-based detector simulation, and the resulting contribution is 
subtracted from the observed yields.

	Continuum background is reduced by requiring that the second 
Fox-Wolfram moment~\cite{fox-wolfram}, $R_2<0.6$. The remainder of this
background is estimated using $\upsl4$ data, and is statistically subtracted
from the observed $\upsln$ yields.
The estimate of this background is discussed in Sections~\ref{sec:bkest} and ~\ref{sec:bkexp}.

	Candidate $\jpsi$'s are formed by pairing oppositely charged
electron or muon candidates. These charged track candidates are required 
to have momentum in the range from 0.1 to 5.3 GeV/$c$ and 
have at least 50\% of the maximum number of expected hits in the tracking system. 
%and have a $\chi^2$/dof$<10$ for the track fit. 
We also require these
tracks to be consistent with coming from the interaction point
in three dimensions.
%In addition, since these particles 
%are expected to come 
%from the interaction point, we require the difference in
%the $r-\phi$ impact parameter to be less than 6.5 mm and the
%difference in the ($r-z$) plane from the nominal interaction point 
%to be less than 10 cm.
Electron candidates are additionally required to have a 
shower profile which is consistent with expectations for an electron and 
an energy deposition in the calorimeter, $E_e$, which is compatible with its measured 
momentum, $p_e$, by requiring $0.85<E_e/p_e<1.15$.
For these electron candidates, we correct for radiated 
photons by adding back the momentum of the highest energy photon which lies within
a 5$^{\circ}$ cone of the initial particle direction. 
Muon candidates are formed using charged tracks which penetrate at least 
3 hadronic interaction lengths of iron absorber in the muon chambers~\cite{cleoiii}.

\section{\boldmath Measurements of $\upstojpsix$}
\label{sec:upstojpsix}

\subsection{\boldmath $\jpsi$ Mass Distributions in the $\upsln$ Data}
\label{sec:yields1s}

	Figure~\ref{mass_1s_all} shows the invariant mass distribution of $\jpsi$
candidates for (a) $\jpsimm$ and (b) $\jpsiee$ in the $\upsln$ on-resonance sample 
satisfying all selection criteria. The shaded histograms show the corresponding
distributions for $\upsl4$ continuum data, scaled by a factor of 0.65, which 
accounts for the differences in luminosities and center of mass energies.
The mass distributions from the $\upsln$ data set are fit to the sum
of a linear background and a Gaussian signal whose means and widths
are allowed to float. The fitted peaks have a resolution of 
13.4 MeV/$c^2$ and 14.2 MeV/$c^2$ for the $\jpsimm$ and $\jpsiee$ channels,
respectively. The fitted yields are 399$\pm$25 $\jpsimm$ and 449$\pm$27 $\jpsiee$ 
signals events.

%	These distributions include both signal and continuum background. The
%$\upsln$ signal includes all intermediate final states, {\it i.e.,} $ggg$, $gg\gamma$,
%and $q\bar{q}$. To
%compute a branching fraction for $\upstojpsix$, we segment the data into 
%scaled momentum bins (discussed below), estimate the background contribution 
%using $\upsl4$ data 
%(extrapolated to the $\upsln$) and subtract it from the observed $\upsln$ yield.

\begin{figure}[btp]
\centerline{
\includegraphics[width=3.5in]{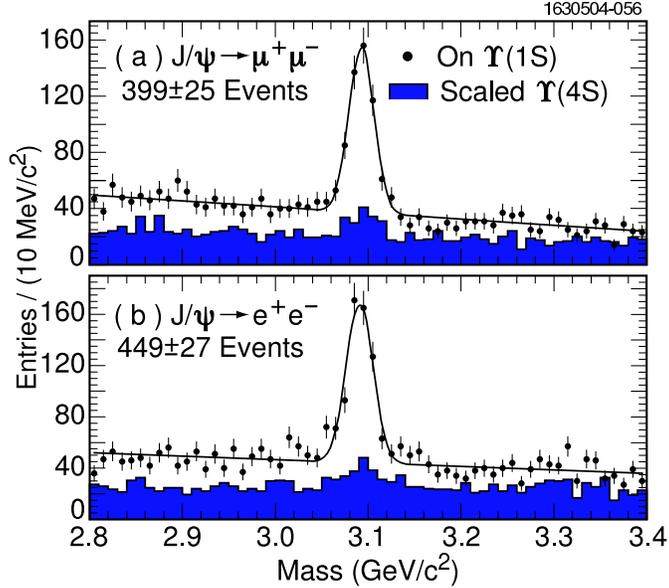}}
\caption{\label{mass_1s_all} 
Dilepton invariant mass distributions for (a) $\jpsi\to\mu^+\mu^-$ and
(b) $\jpsi\to e^+e^-$ candidates for data taken on the $\upsln$ 
resonance (points) and data taken just below the $\upsl4$ 
resonance (shaded). The $\upsl4$ distributions are scaled to 
account for the different integrated luminosities and center 
of mass energies for the two data samples.}
\end{figure}

%\subsection{Momentum Distributions}

	To study the momentum distribution, we divide the data into bins 
of scaled momentum, $x$, where $x=p_{\jpsi}/p_{\rm max}$. Here, 
$p_{\rm max}=(1/2\sqrt{s})(s-M_{\jpsi}^2)$
is the maximum $\jpsi$ momentum assuming the $\jpsi$ is recoiling against
a massless particle, $s$ is the square of the center of mass energy, 
$p_{\jpsi}$ is the momentum of the $\jpsi$ candidate and $M_{\jpsi}$ is 
the $\jpsi$ mass~\cite{pdg}. The data are binned in intervals of 
$\Delta x=0.2$. This scaled momentum variable removes the beam-energy 
dependence which is useful in comparing spectra on the $\upsln$
and the $\upsl4$. The invariant mass distributions for $\jpsimm$ 
and $\jpsiee$ for $\upsln$ data in bins of $x$ are shown in Fig.~\ref{mass_x_mm} and  
Fig.~\ref{mass_x_ee}, respectively. If the $x$ distribution has a sharp peak
near the kinematic endpoint, there may be smearing into the $x>1.0$ region.
The absence of any signal in the $1.0\le x<1.2$ bin shows that all events
are containe within the physically-allowed region.
A simulation of the $\jpsi$ signal (see Section~\ref{sec:receff})
indicates that the widths of the invariant mass 
distributions are independent of $\jpsi$ momentum, and therefore these distributions
are fit using a width fixed to the values obtained from the full sample.

%The efficiency-corrected differential cross-section per unit $x$ is shown in 
%Fig.~\ref{y1s_x_dist}. We have
%also accounted for the $\jpsi$ branching fraction to leptons (5.9\%), and divided 
%through by the integrated luminosity of the $\Upsilon$(1S) data sample (1.2 fb$^{-1}$).
%The distribution is observed to peak at relatively low values of scaled
%momentum, in sharp contrast to predictions of the color-octet model~\cite{cheung_1s},
%but in reasonable agreement with the color-singlet process $\upstojpsiccg$.
%As will be seen in the next section, the background subtraction does not
%affect this observation.

\begin{figure}[btp]
\centerline{
\includegraphics[width=3.5in]{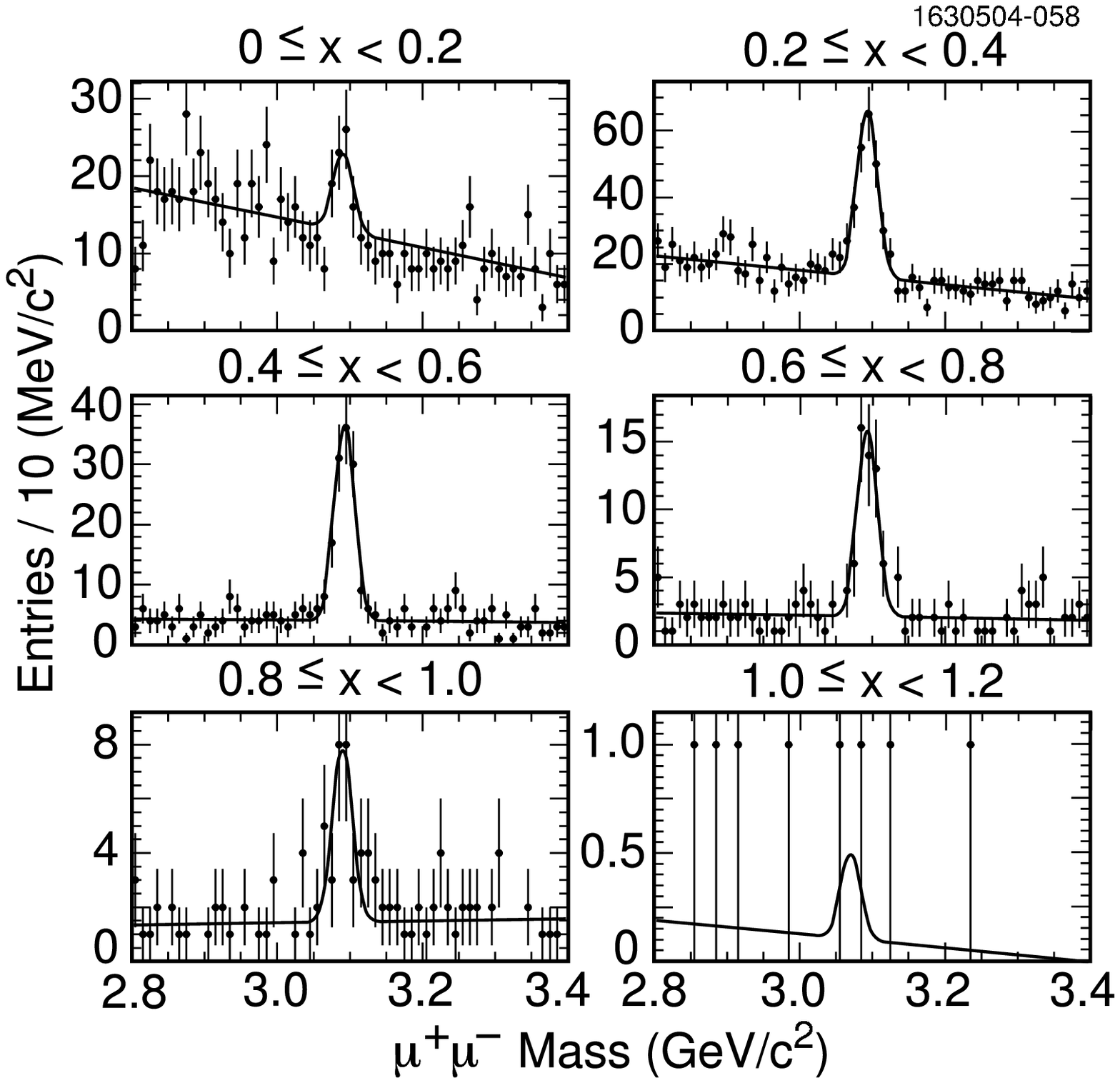}}
\caption{\label{mass_x_mm} 
Invariant mass distributions for $\jpsimm$ candidates in
$x$ bins of size 0.2 for data taken on the $\upsln$ 
resonance.} 
\end{figure}

\begin{figure}[btp]
\centerline{
\includegraphics[width=3.5in]{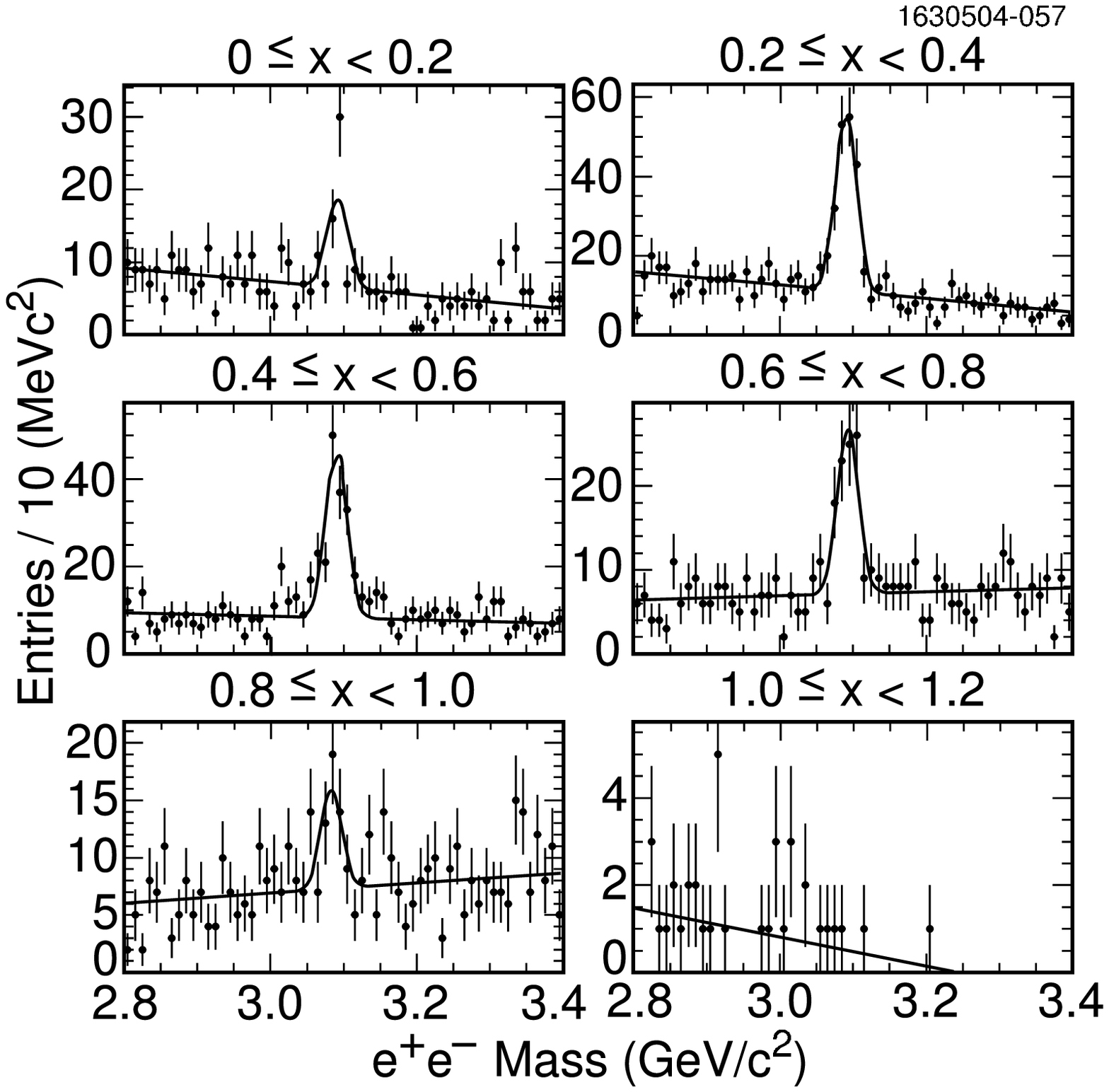}}
\caption{\label{mass_x_ee} 
Invariant mass distributions for $\jpsiee$ candidates in
$x$ bins of size 0.2 for data taken on the $\upsln$
resonance.} 
\end{figure}

\subsection{\boldmath Candidate $\jpsi$ Mass Distributions in the $\upsl4$ Data}
\label{sec:bkest}

	The continuum contribution to the $\upstojpsix$ signal is
estimated using data taken on and below the $\upsl4$. This measurement 
is interesting in itself in
light of the disagreement in the rates for $\eetojpsix$ measured by 
BaBaR ~\cite{babar_ee_jpsi} and Belle~\cite{belle_ee_jpsi}. 
We employ the same event selection criteria as for the data taken
on the $\upsln$, except that for the on-$\upsl4$ data, we require
the $\jpsi$ to have momentum larger than 2 GeV/$c$, which 
eliminates contributions from $B$-meson decay.

	The measured signal for $\eetojpsix$ below the $\upsl4$ is shown in
Fig.~\ref{y4scont_jpsix_tot} for (a) $\jpsimm$ and (b) $\jpsiee$. The
fitted numbers of events are 112$\pm$17 ($\jpsimm$) and 116$\pm$19 ($\jpsiee$).
The corresponding distributions for data taken on the $\upsl4$ resonance
are shown in Fig.~\ref{y4s_p2_jpsix_tot}. The fitted yields are 130$\pm$17
$\jpsimm$ and 193$\pm$24 $\jpsiee$ events. The yields per unit
luminosity are statistically compatible, after correcting the on-$\upsl4$
yield for the 2 GeV/$c$ momentum requirement. The correction is determined
from the $\jpsi$ momentum spectrum from the below-$\upsl4$ continuum and is
estimated to be $(25\pm6)\%$.

\begin{figure}[btp]
\centerline{
\includegraphics[width=3.5in]{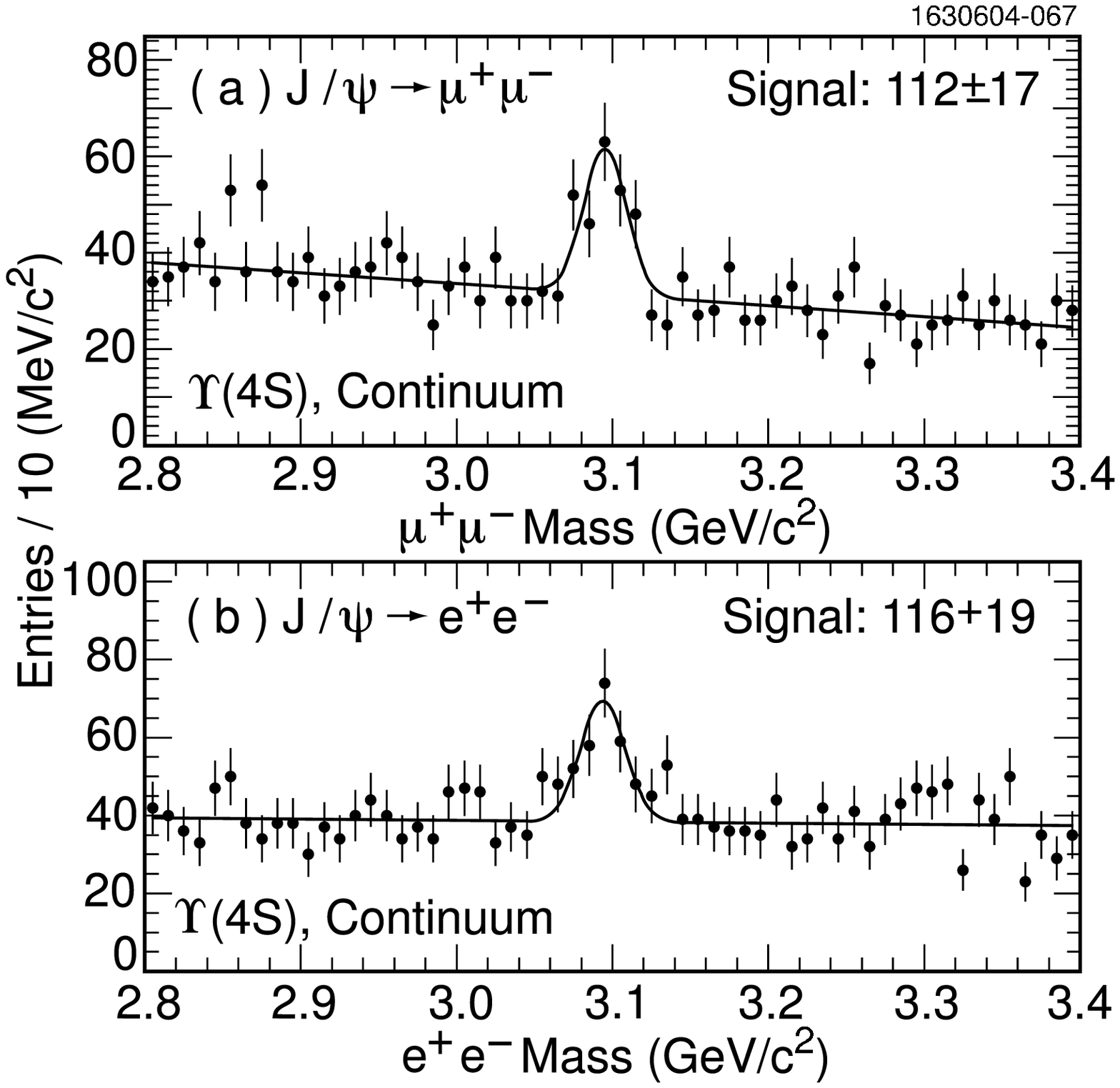}}
\caption{\label{y4scont_jpsix_tot} 
Invariant mass distribution for (a) $\jpsimm$ and (b) $\jpsiee$ in
data taken below the $\upsl4$. The data are integrated over all momenta.}
\end{figure}

\begin{figure}[btp]
\centerline{
\includegraphics[width=3.5in]{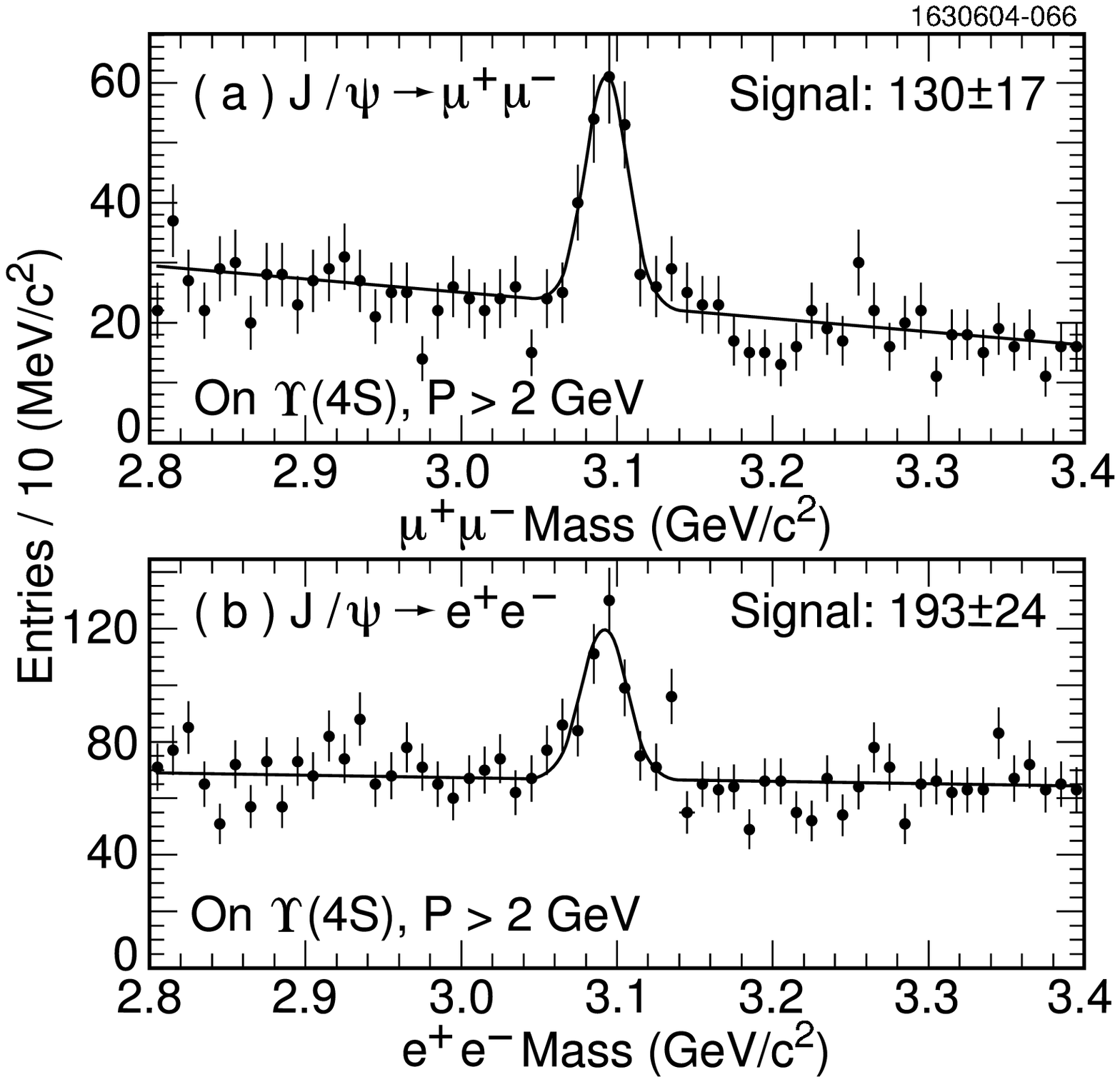}}
\caption{\label{y4s_p2_jpsix_tot} 
Invariant mass distributions for (a) $\jpsimm$ and (b) $\jpsiee$ in
data taken on the $\upsl4$ resonance. To reject $\jpsi$'s from $B$ decay, we 
require the momentum of the $\jpsi$ to be larger than 2.0 GeV/$c$.}
\end{figure}

\subsection{\boldmath $\jpsi$ Reconstruction Efficiency}
\label{sec:receff}

	The data are corrected for geometric acceptance and analysis
requirements using the {\sc pythia} Monte Carlo~\cite{pythia} and a 
{\sc geant}-based detector simulation~\cite{geant}. 
%The PYTHIA Monte Carlo almost exclusively produces $\jpsi$'s in $\upsln$ decays
%in association with a second pair of charmed particles, and (98$\pm$0.3)\%
%of the time, these other two charm quarks produce $D\bar{D}$. The event
%selection requirements are $(98\pm0.5)\%$ efficient for these simulated
%events. However, an issue arises here in that the momentum of the $\jpsi$'s
%in this sample do not exceed about 3.3 GeV/$c$. We therefore 
%supplement the full event simulation with single particle MC simulation 
%using $\psitwos\to\jpsi\pi^+\pi^-$. For this Monte Carlo sample, the 
%track multiplicity and radiative return selection requirements are 
%bypassed. The over-estimated efficiency in this case is less than 2\% 
%which is negligible compared to other uncertainties, and is therefore neglected.
%To check that the efficiencies in the single-$\psitwos$ simulation are not
%over-estimated, the $\jpsi$ momentum region is chosen to overlap that of the
%full event simulation. The two simulations are found to agree quite well
%in the overlap region. We therefore conclude that the efficiency for 
%reconstructing $\jpsimm$ and $\jpsiee$ is not sensitive to moderate event
%multiplicities. 
%As a cross-check, we compared several event-level 
%features such as the number of charged tracks, the neutral energy and $R_2$, 
%between the PYTHIA simulation and $\upstojpsix$ data and we find they agree 
%quite well. 

	The reconstruction efficiency as a function of 
$x$ and $\cos\theta_{\jpsi}$, where $\theta_{\jpsi}$ is the 
polar angle of the $\jpsi$ in the lab frame,
is shown in Fig.~\ref{upstojpsi_receff}. The circular points are 
for $\jpsimm$ and the triangles are for $\jpsiee$.
The efficiencies decrease slightly with increasing momentum
and $|\cos\theta_{\jpsi}|$ and have average values of 
(40$\pm$2)\% for $\jpsimm$ and (50$\pm$2)\% for $\jpsiee$.
The small drop in efficiency with momentum is a result of not
reconstructing the softer lepton which is emitted backward in the
$\jpsi$ rest frame. The lower $\jpsimm$ reconstruction
efficiency is due to the requirement that both 
muons penetrate at least three layers of iron absorber, which
limits the muon momentum to be larger than about 1 GeV/$c$. 

	The momentum distributions of both the $\upstojpsix$ signal
as well as the on-$\upsl4$ and below-$\upsl4$ yields are corrected using these 
$x$-dependent efficiencies. This is justified since the
reconstruction efficiency is not sensitive to small differences in
the event environment (between $\upstojpsix$ and $\eetojpsix$).
Continuum-produced $\jpsi$'s have a similar charged-track 
multiplicity to $\upstojpsix$ ($\approx$7-8) and the $\upsln$ data 
peak at low $R_2$ (see Section~\ref{sec:evlev_dists}, and figures therein)
as do $R_2$ measurements in
the continuum (see Fig.1 (c),(d) in Ref.~\cite{babar_ee_jpsi}).

\begin{figure}[btp]
\begin{center}
\includegraphics[width=3.5in]{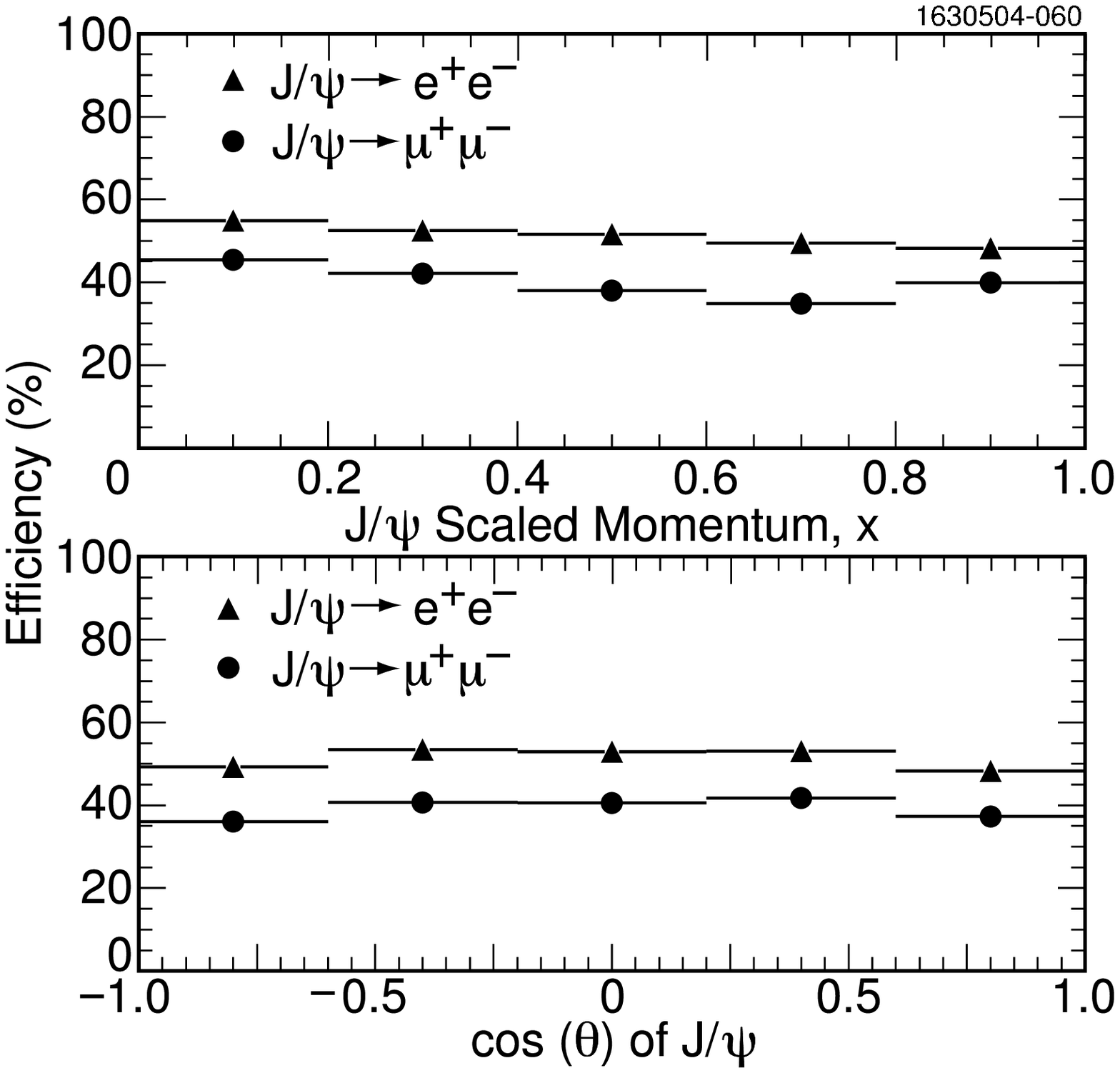}
\caption{\label{upstojpsi_receff} 
Efficiency for reconstructing $\jpsi$'s in $\upsln$ decays as
a function of (a) scaled $\jpsi$ momentum, and (b) $\cos\theta_{\jpsi}$.
The circles are for $\jpsimm$ and the triangles are for $\jpsiee$.}
\end{center}
\end{figure}

\subsection{\boldmath Corrected $\jpsi$ Momentum Distributions on and just below the $\upsl4$ Resonance}
\label{sec:bkspec}

	The resulting differential cross sections, $d\sigma/dx$, versus $x$, 
are shown in Fig.~\ref{y4s_x_dist} using the combined 
on-$\upsl4$ and below-$\upsl4$ data. The circles represent $\jpsimm$ 
and the triangles are $\jpsiee$. The distributions clearly
peak at large $x$ values with a mean of about 0.7.
Integrating these distributions, and using 
${\cal{B}}(\jpsi\to l^+l^-)$=5.9\%~\cite{pdg}, we 
find $\sigma(\eetojpsix)=2.0\pm0.2$(stat) pb for $\jpsimm$ and 
$\sigma(\eetojpsix)=1.7\pm0.2$(stat) pb for $\jpsiee$. Combining
these results we obtain $\sigma(\eetojpsix)=1.9\pm0.2$(stat) pb.
%Using only
%the continuum data, we find cross-sections of 
%$\sigma(\eetojpsix)=2.4\pm0.3$(stat) pb using the $\jpsimm$ channel and
%$\sigma(\eetojpsix)=1.8\pm0.3$(stat) pb using the $\jpsiee$ channel. 
The results using the different lepton species are consistent with one
another and lie between the BaBaR and Belle measurements of
($2.52\pm0.21\pm0.21$) pb and (1.47$\pm$0.10$\pm$0.13) pb, respectively.
The rates found in the continuum are about a factor of 6-7 lower than
on the $\upsln$.

\begin{figure}[btp]
\centerline{
\includegraphics[width=3.5in]{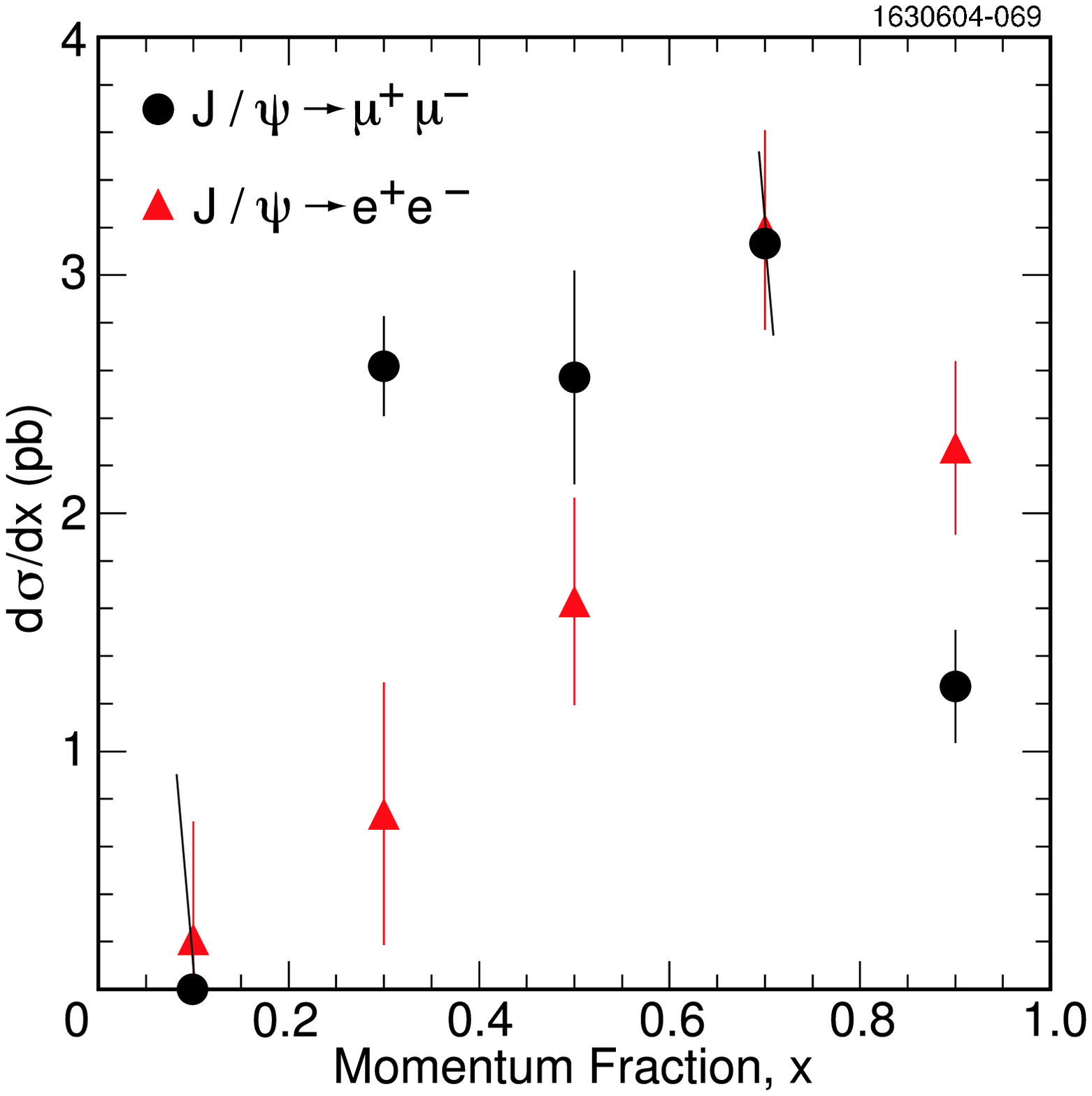}}
\caption{\label{y4s_x_dist}
Distributions in $x$ for $\eetojpsix$ using the combined data taken 
on and just below the $\upsl4$. The circles 
show the results obtained using $\jpsimm$ and the 
triangles show the corresponding distribution obtained using $\jpsiee$.}
\end{figure}

\subsection{\boldmath Extrapolation of the $\upsl4$ results to the $\upsln$}
\label{sec:bkexp}

	The extrapolation of the differential cross section for 
$\eetojpsix$ on and below the $\upsl4$ (see Fig.~\ref{y4s_x_dist}) to the
$\upsln$ requires that we take into account the differences between these
two samples, and includes two factors (other than the luminosity scaling): 
the ratio of partonic cross sections for $\eetojpsix$ and a phase space
correction for producing the $\jpsi+X$ final state. For the former,
we assume $1/s$ scaling, since the process proceeds through
a virtual photon, and therefore the parton-level cross section at 9.46 GeV 
is 1.25 times larger than at 10.58 GeV. For the phase space extrapolation, 
we bound this factor at unity by assuming the phase space at
9.46 GeV is equal to that at 10.58 GeV. To obtain a lower bound, we assume
that the $\jpsi$'s are always produced in association with a pair of $D$ mesons,
which has a significantly reduced phase space at 9.46
GeV as compared to 10.58 GeV. Using {\sc pythia}, we determine that the probability,
of producing $\jpsi D\bar{D}$ at 9.46 GeV is 55\% of the corresponding 
value at 10.58 GeV. Using these values as extremes, and assuming that the 
``true'' value has a flat probability of lying somewhere in that interval,
we estimate the phase space ratio is $0.78\pm0.13$. Combining the two
factors, we determine the continuum extrapolation factor, 
$f_{\rm cont}=0.98\pm0.16$.

	For the $\jpsi$ momentum spectrum in $\upsln$ decays, we are primarily 
interested in the shape for the gluonic intermediate states. The $q\bar{q}$ 
intermediate state, which proceeds through the coupling of the
$\upsln$ to a virtual photon, is assumed to have the same shape in $x$ as in 
$\eetojpsix$, and therefore is more closely related to the predictions
for $\jpsi$ production in the continuum. Therefore for the purposes of the
momentum spectrum, we subtract the expected 
$\upsln\to\gamma^*\to q\bar{q}\to\jpsix$ contribution. 
This contribution is included for the branching fraction measurement.
Any potential interference between the continuum and the 
$\upsln\to\gamma^*\to\ q\bar{q}$ contributions is neglected.
We express the $\upsln\to\gamma^*\to\ q\bar{q}$ rate relative to the 
corresponding rate for $\eetojpsix$ using,

\begin{equation}
{\sigma_{\upstoqq}\over {\sigma_{e^+e^-\to q\bar{q}}} } = 
{ R\times\sigma_{\upstomm}\over {R\times\sigma_{\eetomm}} } = 
{ \sigma_{\upstomm}\over {\sigma_{\eetomm}}. }
\end{equation}

\noindent Here, $\sigma_{\upsln\to X}$ is shorthand for 
$\sigma(e^+e^-\to\upsln)\times {\cal{B}}(\upsln\to X)$.
The measured value for $\sigma_{\upstomm}$ is 0.555$\pm$0.022 nb
~\cite{shaw}. In that same reference, the theoretical value for 
$\sigma_{\eetomm}$ at 9.46 GeV is estimated to be 1.12 nb~\cite{shaw}. A more
recent estimate based on the FPAIR Monte Carlo (MC) simulation~\cite{kleiss} 
gives a larger cross section of about 1.38 nb. Taking the average
of these two cross sections as our central value and half their 
difference as the uncertainty, we obtain $\sigma_{\eetomm}$=1.25$\pm$0.13 nb.
We therefore estimate that the  $\upsln\to\gamma^*\to q\bar{q}\to\jpsix$
contribution is $(44\pm5)\%$ of the $\eetojpsix$ at $\sqrt{s}$=9.46 GeV. 
Adding this contribution to
the continuum extrapolation factor, $f_{\rm cont}$, we obtain
an overall extrapolation factor for the $x$ spectrum of $f_x=1.41\pm0.18$. 

\subsection{\boldmath Corrected $\jpsi$ Momentum Distributions and Branching Fractions in $\upsln$ Data}

	Figure~\ref{y1s_y4s_x_dist} shows the differential cross sections
in $x$ for (a) $\jpsimm$ and (b) $\jpsiee$ using data taken on the 
$\upsln$ (solid circles) and averaged results from
the data taken on and below the $\upsl4$ (triangles). The latter have
been scaled as discussed above to include both the continuum and
$\upsln\to\gamma^*\to q\bar{q}$ contributions. 
The differential cross section (versus $x$) for $\upstojpsix$ is given by the 
difference of these two distributions, and reflects only the contributions from
gluonic intermediate states. 
The results are shown in Fig.~\ref{y1s_sub_x_dist} using $\jpsimm$ (circles)
and $\jpsiee$ (triangles). The figure also shows the theoretical 
predictions of the
color-octet~\cite{cheung_1s} (solid line) and the color-singlet 
$\upsln\to\jpsi+c\bar{c}g$~\cite{li-xie-wang} (dashed line) model. 
%Our data are clearly
%much softer than the color-octet prediction and moderately softer than
%expected from color-singlet production.

%Whether the same 
%softening effects which have been incorporated into the color-octet predictions in 
%$\eetojpsix$~\cite{fleming1} can describe the $\upstojpsix$ spectrum 
%still remains an open issue. 

\begin{figure}[btp]
\centerline{
\includegraphics[width=3.5in]{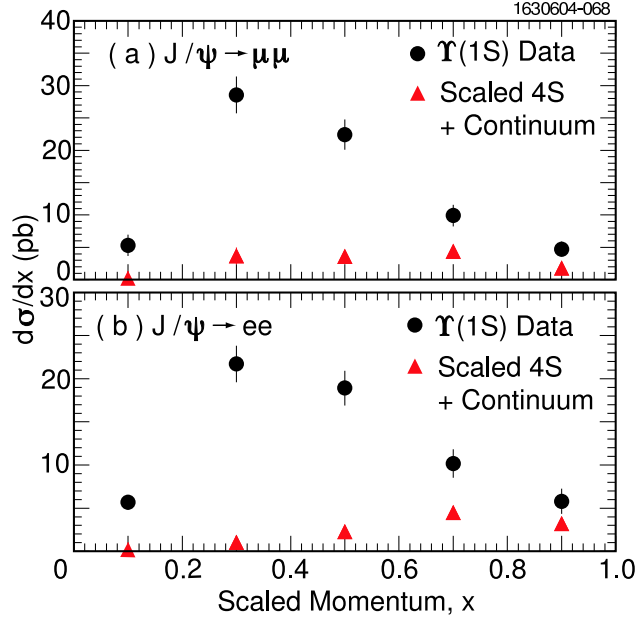}}
\caption{\label{y1s_y4s_x_dist}
Differential cross sections in $x$ for data taken on the $\upsln$
(points) and data taken both on and just below the $\upsl4$ (triangles). 
The latter have been scaled to account for the $\upsln\to\gamma^*\to q\bar{q}$
contribution, as discussed in the text.
The upper figure shows the results obtained using $\jpsimm$ and the lower 
figure shows the corresponding distributions obtained using $\jpsiee$.}
\end{figure}

\begin{figure}[btp]
\centerline{
\includegraphics[width=3.5in]{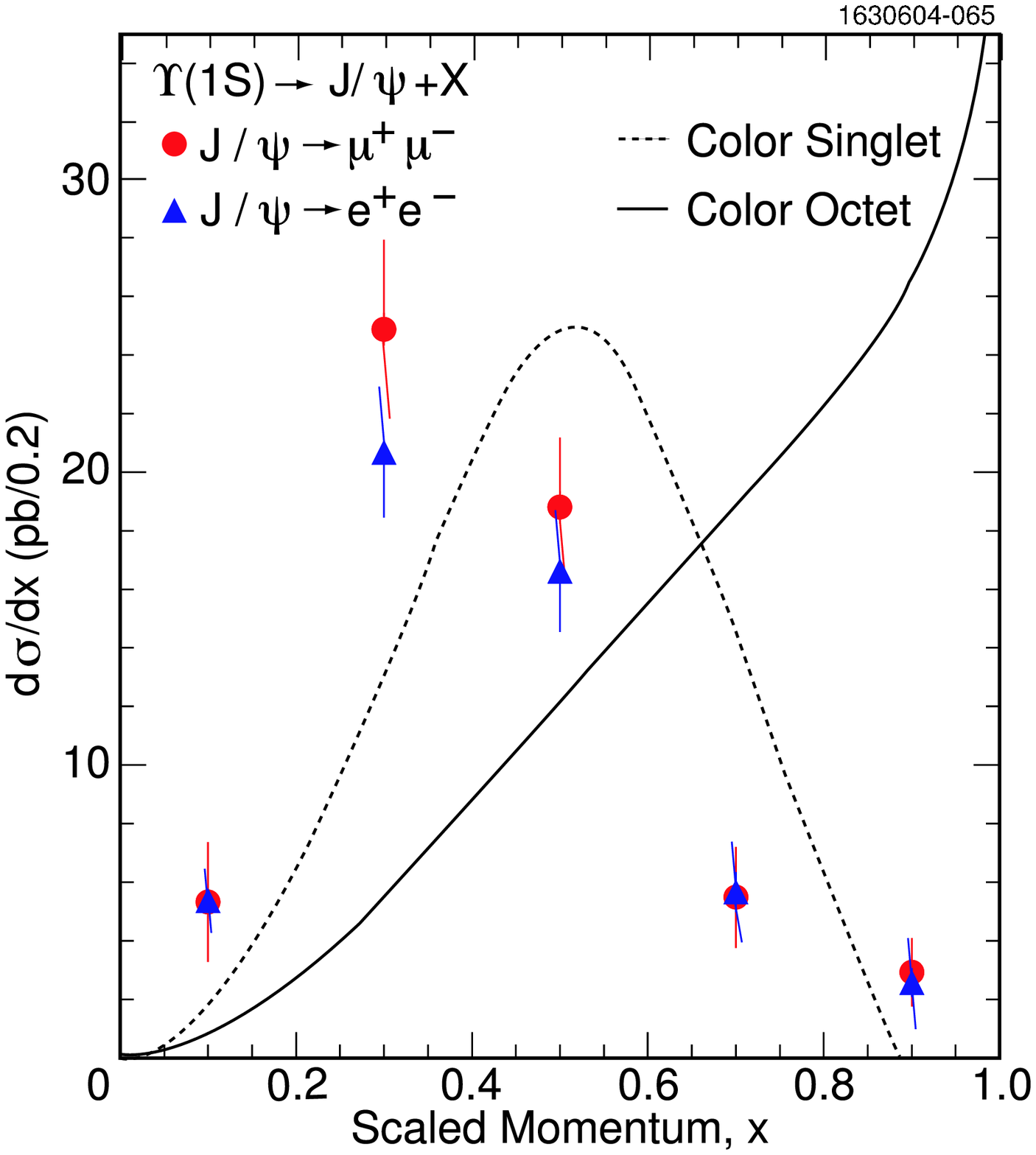}}
\caption{\label{y1s_sub_x_dist}
Differential cross sections in $x$ for $\upstojpsix$ obtained using
$\jpsimm$ (circles) and $\jpsiee$ (triangles). We also show the theoretical
expectations based on the color-octet (solid line)~\cite{cheung_1s} and 
color-singlet (dashed line)~\cite{li-xie-wang} models.}
\end{figure}

	The branching fraction for $\upstojpsix$ is computed by integrating
the differential cross section distribution. We only
subtract the expected continuum contribution (we use $f_{\rm cont}$ as our extrapolation
factor as opposed to $f_x$), so that the branching fraction
includes the three intermediate hadronic states: $ggg$, $gg\gamma^{(\ast)}$ and $q\bar{q}$.
The resulting branching fractions for the $\jpsimm$ and $\jpsiee$ final states are:

\begin{eqnarray}
{\cal{B}_{\mu\mu}}(\upstojpsix) = (6.9\pm0.5({\rm stat}))\times 10^{-4}  \nonumber \\
{\cal{B}}_{ee}(\upstojpsix) = (6.1\pm0.5({\rm stat}))\times 10^{-4}.
\end{eqnarray}

\noindent Systematic uncertainties are presented in Section~\ref{sec:sys}.
Using $\Gamma_{\rm tot}(\upsln)=53.0\pm1.5$)~keV~\cite{pdg}, 
our measurement corresponds to partial widths, $\Gamma_{ggg+gg\gamma+q\bar{q}}$ 
of ($36.6\pm2.8$) eV and ($32.3\pm2.8$) eV
for the $\jpsimm$ and $\jpsiee$ channels, respectively.

	Subtracting the expected  $\upsln\to\gamma^*\to q\bar{q}$ 
contribution, we obtain  $\Gamma_{ggg+gg\gamma^{(\ast)}}$ of  
($33.9\pm2.8$) eV and ($30.2\pm2.3$) eV. In other words, about 90\% of the
$\jpsi$ rate comes from the $ggg$ and $gg\gamma^{(\ast)}$ intermediate states. 
The $gg\gamma^{(\ast)}$ contribution is only expected to be at the level of about 5\%~\cite{cheung_1s} of the $ggg$ rate.

	Theoretical estimates of this rate based only on 
color-octet contributions, which neglect the $q\bar{q}$ intermediate state,
give a total branching fraction of $6.2\times 10^{-4}$~\cite{cheung_1s,napsuciale}. 
Those predictions are in good agreement with the measurements reported here.
On the other hand, our measured momentum spectrum is significantly softer than predicted 
by the color-octet model, which is expected to peak near the kinematic limit
(see Fig.~\ref{y1s_sub_x_dist}). However, it has been recently pointed out~\cite{fleming1}
that in a similar process, $\eetojpsix$, the non-relativistic 
calculations break down near the kinematic endpoint where there are large perturbative
and non-perturbative corrections. These effects may be systematically treated using
so-called Soft-Collinear Effecive Theory (SCET) and are expected to soften the 
$\jpsi$ momentum spectrum. Using SCET, the shape of the measured $\jpsi$ momentum spectrum 
in $\eetojpsix$, which peaks near $x\simeq 0.7$, was shown to be reproducible~\cite{fleming1}. 
It will be interesting to see if these corrections, when applied to $\upstojpsix$, can soften the
color-octet predictions sufficiently to bring them into agreement with our data.

	Our measured rate is also consistent with
the predictions of the color-singlet process $\upstojpsiccg$, which predicts
a branching fraction of $5.9\times 10^{-4}$ and a soft momentum spectrum
which peaks at $x\approx0.5$ and has a kinematic limit of 
$x<0.9$. While the data is somewhat softer than the color-singlet 
predictions, it should be noted that this is a parton-level calculation and neglects
the hadronization process. Inclusion of the hadronization of the charm quarks 
to charm hadrons softens the $\jpsi$ momentum spectrum,
with more softenining occuring as the mass of the recoiling system increases.
Further softening of the $\jpsi$ momentum spectrum occurs when including the
feed-down of $\psitwos$ and $\chi_{cJ}$ to $\jpsi$. 
Using a {\sc pythia} simulation of the color-singlet process, 
we are able to obtain reasonably good agreement in the $x<0.6$ region using 
our measured values for the feed-down from $\psitwos$, $\chi_{cJ}$ to $\jpsi$ 
along with a reasonable, but arbitrary admixture of recoiling 
$D$, $D^*$, and $D^{**}$ states. This is not necessarily evidence for color-singlet
production, but it is suggestive.

\subsection{\boldmath $\jpsi$ Angular Distributions in $\upsln$ Data}

	Angular distributions have the potential to differentiate the
mechanisms for $\jpsi$ production in $e^+e^-$ collisions.
Theoretical predictions for the production and helicity angle distributions
for continuum production are available~\cite{ee_jpsi,ang_theory1,ang_theory2}, 
but the calculations are yet to be done for $\upsln$ decay.

	In the same spirit, we present distributions of the (polar) production angle,
$\cos\theta_{\jpsi}$, of the $\jpsi$ and the helicity angle, $\cos\theta_{\rm hel}$, 
where $\theta_{\rm hel}$ is the angle between the positive lepton momentum in the $\jpsi$
rest frame and the $\jpsi$ momentum in the lab frame. The efficiency-corrected $\jpsimm$ and
$\jpsiee$ channels are combined and shown in Fig.~\ref{y1s_ang_dists}. Here,
we subtract the expected $\upsln\to\gamma^*\to q\bar{q}$ and continuum 
contributions to extract the $ggg$ and $gg\gamma^{(\ast)}$ shapes. The normalizations
are arbitrary. The top
figure shows the distribution of $\cos\theta_{\rm hel}$ for $\upsln$ data (points),
{\sc pythia} simulation (dashed), and a fit (dotted) to the form $(1+A\cos^2\theta_{\rm hel})$,
from which we find $A=-0.48\pm0.16$ ($\chi^2$/dof=0.50). The bottom figure shows the distribution in 
$\cos\theta_{\jpsi}$ for $\upsln$ data (points), {\sc pythia} simulation (dashed) and a 
fit (dotted) to the form $(1+B\cos^2\theta_{\jpsi})$, from which we find $B=0.01\pm0.16$ ($\chi^2$/dof=1.44). 
The functional forms are the same as those used to describe the angular distributions for
continuum production of $\jpsi$ mesons~\cite{ee_jpsi,ang_theory2}.
The negative value of $A$ indicates that the $\jpsi$ has a significant longitudinal
polarization component (a positive value would indicate transverse polarization). 
For continuum production of $\jpsi$, the color-octet and color-singlet models differ greatly 
on their expectations for $B$ at large values of scaled momentum, with $B\simeq-0.85$
for the color-singlet model~\cite{ang_theory2} and $B\approx +1$ for the color-octet 
model~\cite{ee_jpsi}. If a large difference persists for $\upsln$ decay, the production
angle distribution could be useful in differentiating these two mechanisms. We note
that the {\sc pythia} simulation (using default parameters), which produces $\jpsi$ via
$\upsln\to\jpsi+D\bar{D}$, appears to be in reasonable agreement with data.

\begin{figure}[btp]
\centerline{
\includegraphics[width=3.5in]{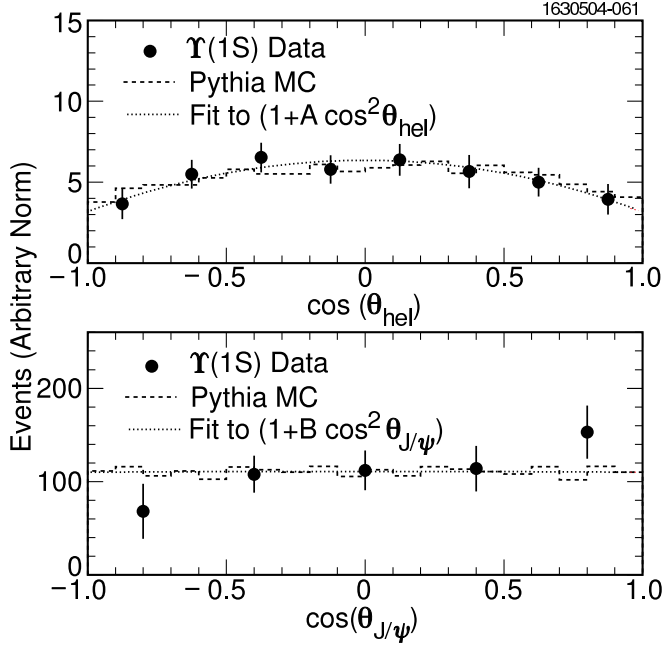}}
\caption{\label{y1s_ang_dists}
The helicity angular distributions, $\cos\theta_{\rm hel}$ (top) and production
angle, $\cos\theta_{\jpsi}$ (bottom) of the $\jpsi$ in $\upstojpsix$. In each 
case, the points are the $\upsln$ data, the dashed histogram is a {\sc pythia} 
simulation of $\upstojpsix$ and the dotted line
is a fit to the $\upsln$ data as described in the text.}
\end{figure}

\subsection{\boldmath Event-Level Distributions}
\label{sec:evlev_dists}

	Additional information on the $\upstojpsix$ process can be obtained by
studying various event-level distributions. We present distributions of:

\begin{itemize}
 \item Number of reconstructed charged tracks, $N_{\rm trk}$ (Fig.~\ref{y1s_dists}(a))
 \item Reconstructed neutral energy in the crystal calorimeter, $E_{\rm NEU}$  (Fig.~\ref{y1s_dists}(b))
 \item The second Fox-Wolfram moment, $R_2$ (Fig.~\ref{y1s_dists}(c))
 \item Invariant mass recoiling against the $\jpsi$, $M_{\rm RECOIL}$ (Fig.~\ref{y1s_dists}(d))
\end{itemize}

\noindent 
In each case, we have performed a sideband subtraction, where the signal
region is defined to be from $3.04<M_{l^+l^-}<3.14$ GeV/$c^2$ 
and the sideband region includes the $M_{l^+l^-}$ regions from $2.90-2.95$ and
$3.20-3.25$ GeV/$c^2$. The relatively small continuum contribution has not
been subtracted. For each distribution, we also show the corresponding
distribution from the {\sc pythia} MC simulation, 
which primarily produces a final state which consists of $\jpsi D\bar{D}$. 
The data are shown as points ($\jpsimm$ are circles and $\jpsiee$ are triangles)
and the simulation is the histogram. 
We find that the charged particle multiplicity, which includes all
charged particles, has a mean of about nine.
The neutral energy, which comprises all energy in the calorimeter
which is not associated with charged tracks, has an average
of about ($1.5-2.0$) GeV, with most of the
events having less than about 3.5 GeV. The Fox-Wolfram moment, $R_2$,
peaks at low $R_2$ in
$\upstojpsix$ data which indicates that these events tend to be more spherical
than collimated (jetlike).
The recoil mass can be used to discern whether there is another particle
recoiling against the $\jpsi$. It is defined by:

\begin{equation}
M_{\rm RECOIL} = \sqrt{(\sqrt{s}-E_{\jpsi})^2 - p_{\jpsi}^2},
\end{equation}

\noindent where $s$ is the square of the center-of-mass energy, and $p_{\jpsi}$ and
$E_{\jpsi}$ are the momentum and energy of the $\jpsi$ candidate. 
We do not observe any significant peaks in the recoil mass spectrum, 
indicating that the $\jpsi$ is usually not accompanied only by a second 
(bound) $c\bar{c}$ meson. The color-octet model predicts $\sim$1\%
contribution to the inclusive rate whereas the color-singlet model does
not predict the fraction of recoiling charm which is in the form of charmonium.

\begin{figure}[btp]
\centerline{
\includegraphics[width=3.5in]{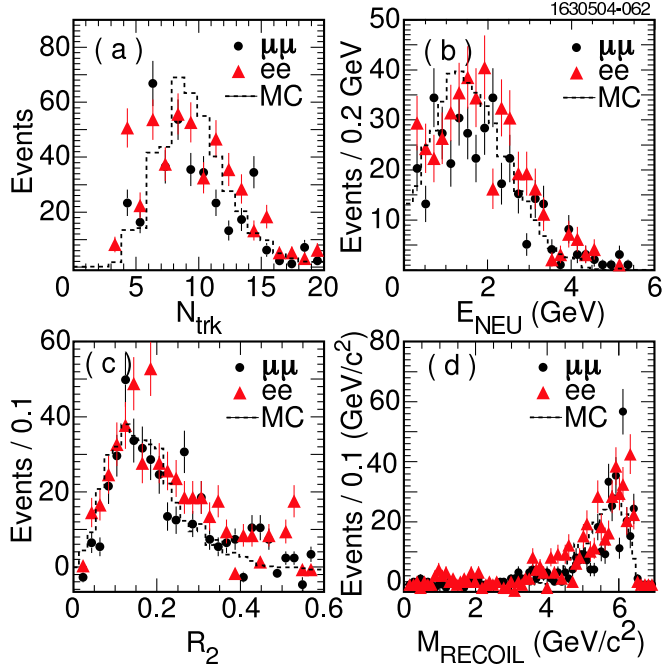}}
\caption{\label{y1s_dists} 
Sideband-subtracted distributions of (a) number of charged tracks, $N_{\rm trk}$,
(b) neutral energy, $E_{\rm NEU}$, (c) the second Fox-Wolfram moment, $R_2$, and 
(d) the recoil mass, $M_{\rm RECOIL}$,
for $\upsln$ data and {\sc pythia} simulation. The points (triangles) correspond to
$\jpsimm$ ($\jpsiee$), and the histograms are the corresponding distributions
from the simulation.}
\end{figure}

\subsection{\boldmath Cross-check using $B\to\jpsi+X$\label{sec:bf_b_to_jpsix}}

	As a cross-check of our detector simulation and analysis procedure, 
we use the same tools to measure ${\cal{B}}(B\to\jpsi+X$) in $\upsl4$ data. The efficiencies
for reconstructing $\jpsi$ in $B\to\jpsi+X$ events are about 5\% lower than
in $\upstojpsix$. In addition to the selection requirements described in 
Section~\ref{sec:evsel},
we require the $\jpsi$ momentum to be less than 2.0 GeV/$c$. The yields
are corrected for the expected continuum contribution, which is typically at the level
of 1-2\% of the $B\to\jpsi+X$ yield. The resulting branching fractions
are found to be $(1.17\pm0.03$(stat))\% and $(1.14\pm0.02$(stat))\% for 
$\jpsimm$ and $\jpsiee$, respectively. These results are slightly higher than the 
world average value of $(1.090\pm0.035)$\%~\cite{pdg}. This difference is 
included as a 
systematic uncertainty in the $\jpsi$ reconstruction efficiency.

\section{\boldmath Measurements of $\upsln\to\psitwos+X$}
\label{sec:psitwos}

\subsection{\boldmath Measurements in $\upsln$ Data} 

	We search for $\upsln\to\psitwos+X$ using the decay mode
$\psitwos\to\jpsi\pi^+\pi^-$. Pion candidates must pass
the previously mentioned track selection criteria and must have a measured
energy loss in the tracking chambers within four standard deviations
of the expected value. Using all pairs of oppositely charged pion candidates, 
we compute the
invariant mass difference, $M(l^+l^-\pi^+\pi^-)-M(l^+l^-)$, a quantity which has
better resolution than $M(\jpsi\pi^+\pi^-)$. 
We also require $M(l^+l^-)$ to be in the range from 3.00 - 3.14 GeV/$c^2$.
The resulting distribution for $M(l^+l^-\pi^+\pi^-)-M(l^+l^-)$ 
is shown in Fig.~\ref{ups1s_psi2s_em}, where we have summed over both
lepton species.
The distribution is fit to the sum of a Gaussian signal shape and a
second-order polynomial background. The width of the Gaussian is
fixed to 2.5 MeV/$c^2$, the value determined from $B\to\psitwos+X$ data.
The 0.3 MeV intrinsic width~\cite{pdg} of the $\psitwos$ is negligible 
compared to the detector resolution, and is therefore ignored.
The fitted yield is 56$\pm$11 events. The significance of the signal,
$S/\sqrt{S+B}$, where $S$ is the fitted signal and $B$ is the estimated 
background, varies from $6-7$, depending on whether $B$ is estimated from
the sidebands or the background function. 

\begin{figure}[btp]
\centerline{
\includegraphics[width=3.5in]{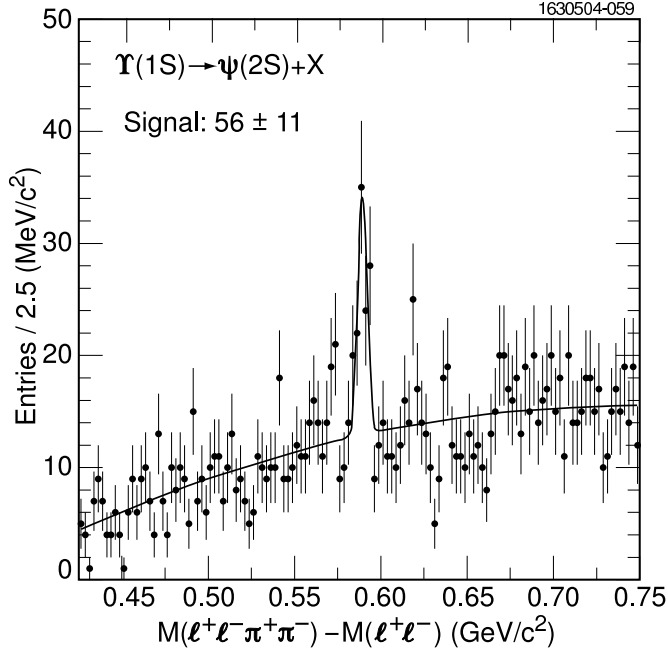}}
\caption{\label{ups1s_psi2s_em} 
Invariant mass difference $M(l^+l^-\pi^+\pi^-)-M(l^+l^-)$ for 
both $\jpsi\to\mu^+\mu^-$ and $\jpsi\to e^+e^-$ candidates with
invariant mass in the range from 3.00-3.14 GeV/$c^2$. }
\end{figure}

	The radiative return background, $e^+e^-\to\psitwos\gamma$, is estimated
using the {\sc evtgen}~\cite{evtgen} simulation package and published 
cross sections in 
Ref.~\cite{benayoun}. The events are processed using {\sc geant} and
analyzed using the same analysis tools as the $\upsln$ data. 
The efficiency for these events to pass a loose hadronic
event selection is (1.4$\pm$0.1)\% for $\jpsimm$ decays
and (8.4$\pm$0.6)\% for $\jpsiee$ decays. For these subsamples, a fraction, 
$f_{\mu\mu}^{\rm radret, pass}=0.40\pm 0.09$ of $\jpsimm$ decays also pass
the analysis-specific selection criteria discussed in Section~\ref{sec:evsel}. 
The corresponding fraction
for $\jpsiee$ decays is $f_{ee}^{\rm radret, pass}=0.15\pm0.02$.
The larger efficiency for the electron channel to pass the loose hadronic
event selection results from the use of the calorimeter in defining
this subsample of events. With the assumption that all data
events which fail the analysis requirements are radiative return
(discussed below), the expected background 
contribution in data from radiative return is computed using

\begin{equation}
N_{ll}^{\rm radret} = {\left(f_{ll}^{\rm radret, pass} 
\over f_{ll}^{\rm radret, rej}\right)}_{\rm MC} 
N_{ll}^{\rm data, rej},
\end{equation}

\noindent where the quantity in parentheses is the ratio of simulated 
radiative return events which pass the analysis-specific selection to those that
are rejected. The quantity $N_{ll}^{\rm data, rej}$ is the number of
rejected events in $\upsln$ data for each lepton species.
We find $N_{\mu\mu}^{\rm data, rej}=5$ and $N_{ee}^{\rm data, rej}=39$ 
in the $\psitwos$ signal region, obtained through sideband subtraction.
We therefore estimate radiative return contributions of 2.0$\pm$1.0 events
and 5.9$\pm$1.0 events in the $\jpsimm$ and $\jpsiee$ channels, respectively,
and therefore a total of 7.9$\pm$1.4 background events from this source.

	The assumption that the rejected events in data are from radiative 
return is 
supported by comparing event-level distributions of these rejected events between
data and simulated $e^+e^-\to\psitwos\gamma$ radiative return events.
Figure~\ref{y1s_psitwos_rej} shows comparisons of (a) number of reconstructed
charged tracks, $N_{\rm trk}$, (b) neutral energy in the calorimeter, $E_{\rm NEU}$,
(c) missing event momentum, $P_{\rm event}$, and (d) the cosine of the angle between
the $\psitwos$ direction and the beam axis, $\cos\theta_{\psitwos}$. In all cases,
the radiative return simulation reproduces the rejected
events in $\upsln$ data, indicating that the rejected data events are mostly
from radiative return.

\begin{figure}[btp]
\centerline{
\includegraphics[width=3.5in]{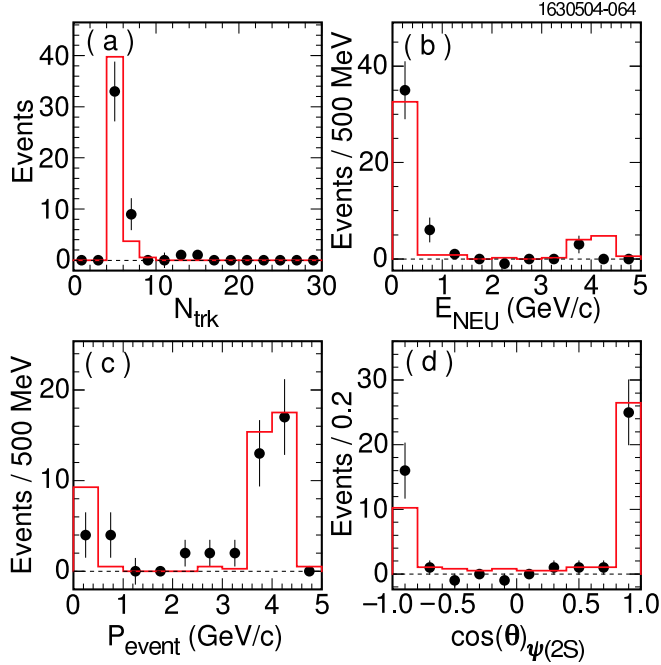}}
\caption{\label{y1s_psitwos_rej}
Event-level distributions for events rejected in the $\psitwos$ analysis.
Distributions shown are (a) number of charged tracks, (b) neutral
energy, (c) missing event momentum, and (d) cosine of the $\psitwos$ production
angle. Solid points are rejected $\upsln\to\psitwos+X$ candidate events  
and the histogram is the $e^+e^-\to\psitwos+X$ radiative return simulation.} 
\end{figure}

	In Fig.~\ref{y1s_psitwos_acc} we show the analogous distributions
for $\upsln$ data (points) passing all analysis selection requirements. The 
corresponding distributions from a MC simulation of 
$\psitwos D\bar{D}$ are overlaid (histogram). These distributions
are clearly quite different than the distributions for rejected events
(see Fig.~\ref{y1s_psitwos_rej}).

\begin{figure}[btp]
\centerline{
\includegraphics[width=3.5in]{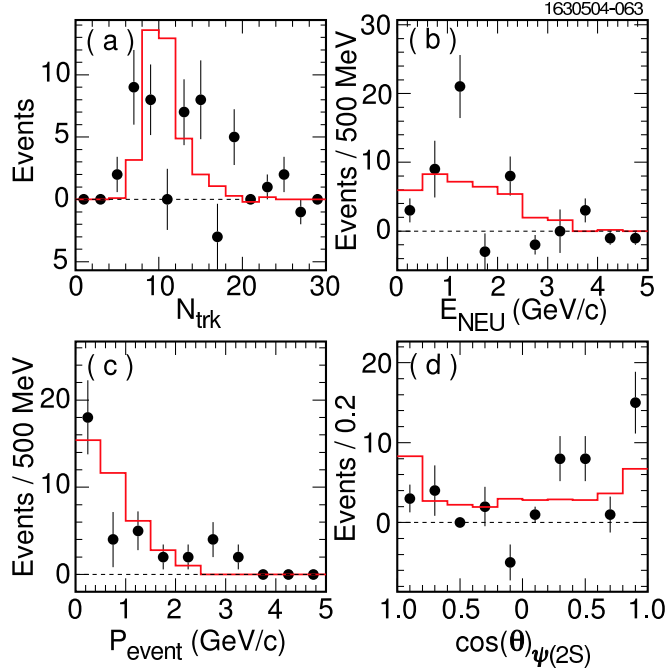}}
\caption{\label{y1s_psitwos_acc}
Event-level distributions for events accepted in the $\psitwos$ analysis.
Distributions shown are (a) number of charged tracks, (b) neutral
energy, (c) missing event momentum, and (d) cosine of the $\psitwos$ production
angle. Solid points are accepted $\upsln\to\psitwos+X$ candidate events  
and the histogram is a $\upsln\to\psitwos D\bar{D}$ MC simulation.} 
\end{figure}

	The continuum background contribution is estimated using
the measured cross section, 
$\sigma(e^+e^-\to\psitwos+X)=0.67\pm0.09^{+0.09}_{-0.11}$,
by Belle~\cite{belle_ee_jpsi}. The expected number of $\psitwos$
continuum background events is then given by

\begin{widetext}
\begin{equation}
N_{\rm back, \rm exp}^{\psitwos} = \sigma(e^+e^-\to\psitwos+X)\times{\cal{L}}
\times {\cal{B}}(\psitwos\to\jpsi\pi^+\pi^-)\times
{\cal{B}}(\jpsi\to l^+l^-)\times\epsilon^{\psitwos}_{ll}\times f_{\rm cont}.
\label{eq:nrec}
\end{equation}
\end{widetext}

\noindent We use the integrated luminosity ${\cal{L}}=1.2~\ifb$,
and branching ratios of ${\cal{B}}(\psitwos\to\jpsi\pi^+\pi^-)=0.318\pm0.010$~\cite{pdg} 
and ${\cal{B}}(\jpsi\to l^+l^-)=0.059\pm0.001$. 
The reconstruction efficiencies are determined using a 
{\sc pythia} simulation of $\upsln\to\psitwos+X$ which is used to
model the continuum as well as the signal, for the same reasons as mentioned
previously. The efficiencies for both $\jpsimm$ and $\jpsiee$ final states
are nearly independent of momentum with average values,
$\epsilon^{\psitwos}_{ll}$=(17$\pm$1)\% for $l=\mu$ and 
($23\pm 1)$\% for $l=e$.
For the background extrapolation, we have assumed 
the same phase space suppression for $\psitwos$ as $\jpsi$, and
assign a 50\% uncertainty to its value.
We therefore expect a continuum background contribution of 
2.5$\pm$1.3 $\jpsimm$ and 3.4$\pm$1.8 $\jpsiee$ events, which sum to
5.9$\pm$2.2 events. The error is dominated by the uncertainties in 
$\sigma(e^+e^-\to\psitwos+X)$ and $f_{\rm cont}$. As a 
consistency check, we have searched our 2.3 $\ifb$ continuum data sample
for $\psitwos$, and we find $2.6^{+4.0}_{-2.6}$ and 12$\pm$4 events
in the $\mu\mu$ and $ee$ channels, respectively. Using the Belle cross section
measurement, we would have expected $4.8\pm0.7$ $\mu\mu$ and $6.5\pm1.0$ $ee$ events,
which is consistent with our observations. 

	Combining the radiative return and continuum backgrounds, we 
estimate a total background of 13.8$\pm$2.6 events. The uncertainty
in the central value is included as a systematic uncertainty (see
Section~\ref{sec:sys}).

	We now compute ${\cal{B}}(\upsln\to\psitwos+X)$.
To reduce systematic uncertainty, the $\psitwos$ branching 
fraction is computed relative to ${\cal{B}}(\upstojpsix)$
and is given by

\begin{widetext}
\begin{equation}
{{\cal{B}}(\upsln\to\psitwos+~X)\over {\cal{B}}(\upstojpsix)} = 
{1\over {\cal{B}}(\psitwos\to\jpsi\pi^+\pi^-)}
{\left(N_{ll,\rm rec}^{\psitwos}-N_{ll,\rm back}^{\psitwos} \over 
 N_{ll,\rm rec}^{\jpsi}-N_{ll,\rm back}^{\jpsi}\right)}
{\left(\epsilon^{\jpsi}_{ll}\over \epsilon^{\psitwos}_{ll}\right)},
\label{eq:bfeq2s}
\end{equation}
\end{widetext}

\noindent where $N_{ll,\rm rec}^{\psitwos}$ ($N_{ll,\rm rec}^{\jpsi}$)
is the total number of $\psitwos$ ($\jpsi$) signal
candidates for lepton species $l=e$ and $l=\mu$, $N_{ll,\rm back}^{\psitwos}$
($N_{ll,\rm back}^{\jpsi}$) is the expected $\psitwos$ ($\jpsi$) background,
and $\epsilon^{\psitwos}_{ll}$ ($\epsilon^{\jpsi}_{ll}$) is the average
reconstruction efficiency for $\psitwos$ ($\jpsi$).
A summary of the inputs used for the $\upsln\to\psitwos+X$ branching fraction
computation are presented in column 4 of Table~\ref{tab:psi2s}. The table 
also shows in columns 2 and 3 the values for the $\jpsimm$ and $\jpsiee$ 
channels separately. The event yields are consistent with one another.

\begin{table*}[hbt] 
\begin{center}  
\caption{Various quantities relevant to the $\psitwos$ analysis as
discussed in the text. Quantities included are the number of reconstructed
candidates $\jpsi$ and $\psitwos$ candidates, their expected backgrounds,
and efficiencies. The bottom two lines give the computed ratio of branching
fractions, as discussed in the text.
\label{tab:psi2s}.} 
\begin{tabular}{|c|c|c|c|}\hline 
Quantity                     & $\mu\mu$     & $ee$ & Combined ($ee+\mu\mu$)\\
\hline
$N_{ll\rm rec}^{\jpsi}$      & 399$\pm$25   & 449$\pm$27   & 848$\pm$37 \\
$N_{ll\rm back}^{\jpsi}$     & 53$\pm$11    &  66$\pm$13   & 119$\pm$17 \\
$\epsilon_{ll}^{\jpsi}$      & (40$\pm$2)\% & (50$\pm$2)\% & (45$\pm$2)\% \\
$N_{ll\rm rec}^{\psitwos}$   & 21$\pm$7     &  35$\pm$8    & 56$\pm$11 \\
$N_{ll\rm back}^{\psitwos}$  &4.5$\pm$1.6   &  9.3$\pm$2.0 & 13.8$\pm$2.6\\
$\epsilon_{ll}^{\psitwos}$   & (17$\pm$1)\% & (23$\pm$1)\% & (20$\pm$1)\%\\
\hline
${{\cal{B}}(\upsln\to\psitwos+X)\over {\cal{B}}(\upsln\to\jpsi+X)}$ &  
                             0.35$\pm$0.15 & 0.46$\pm$0.15 & 0.41$\pm$0.11 \\
\hline
${{\cal{B}}(\upsln\to\psitwos+X){\cal{B}}(\psitwos\to\jpsi+X)\over {\cal{B}}(\upsln\to\jpsi+X)}$ & 
			     0.20$\pm$0.09 & 0.27$\pm$0.09 & 0.24$\pm$0.06 \\
\hline
\end{tabular} 
\end{center} 
\end{table*} 

	In Table~\ref{tab:psi2s}, the number of $\jpsi$ background events 
in the 
$\upsln$ data is computed using the average measured cross section for 
$e^+e^-\to\jpsi+X$ of ($1.9\pm0.2$) pb, the average efficiencies (also shown in
Table~\ref{tab:psi2s}) and the continuum extrapolation factor $f_{\rm cont}$ 
discussed in Section~\ref{sec:bkexp}. 
The ratio of branching fractions is computed to be:

\begin{equation}
{{\cal{B}}(\upsln\to\psitwos+X)
\over {\cal{B}}(\upsln\to\jpsi+X)} = 0.41\pm0.11({\rm stat}). \nonumber \\
\end{equation}

\noindent That is, we find that the rate for 
$\upsln\to\psitwos+X$ is (41$\pm$11)\% of the rate for
$\upstojpsix$ (systematic uncertainties are discussed in 
Section~\ref{sec:sys}). It is interesting
to note that in the $\upsl4$ continuum, Belle finds this ratio to be about 
0.45~\cite{belle_ee_jpsi} with about a 20\% relative uncertainty. Using 
${\cal{B}}(\psitwos\to\jpsi+X)$=(57.9$\pm$1.9)\%~\cite{pdg}, we 
find the feed-down contribution of $\psitwos$ to $\upstojpsix$ to be:

\begin{widetext}
\begin{equation}
{{\cal{B}}(\upsln\to\psitwos+X){\cal{B}}(\psitwos\to\jpsi+X)
\over {\cal{B}}(\upsln\to\jpsi+X)} = 0.24\pm0.06({\rm stat}). \nonumber \\
\end{equation}
\end{widetext}

\noindent This ratio is significantly higher than the expectations 
of either the color-octet model~\cite{cheung_1s} or the color-singlet model
in Ref.~\cite{li-xie-wang}, each which predict a feed-down rate to be
 about 10\%.

\subsection{\boldmath Cross-check by measuring $B\to\psitwos+X$ in $\upsl4$ Data}

	As a cross-check on our analysis, we measure the yield
for $B\to\psitwos+X$ using 10.4 million $B$-meson decays from the
$\upsl4$ data sample, and use the same simulation tools to 
translate this into a branching fraction. The analysis techniques 
are also identical, except that we additionally require the momentum of 
the $\psitwos$ to be less than 1.5 GeV/$c$, which is the kinematic limit for
its production in $B$-meson decay. We find 129$\pm$16 and 144$\pm$18 signal events in
the $\jpsimm$ and $\jpsiee$ channels, respectively, of which
3$\pm$1 and 4$\pm$1 events are expected from continuum background.
The efficiencies are determined using generated {\sc pythia} $b\bar{b}$ events
followed by a full detector simulation, and they 
are found to be (18$\pm$2)\% for the $\jpsimm$ channel and (24$\pm$2)\%
for the $\jpsiee$ channel. 
The branching fractions are measured to be 
$(3.6\pm0.4({\rm stat}))\times 10^{-3}$
for the $\jpsimm$ channel and $(3.0\pm0.4({\rm stat}))\times 10^{-3}$
for the $\jpsiee$ channel. Thus we obtain good agreement with the world 
average value of $(3.10\pm0.24)\times 10^{-3}$~\cite{pdg}.

\section{\boldmath Measurements of $\chi_{cJ}$ in $\upsln$ and $\upsl4$ Data}
\label{sec:upstochicx}

\subsection{\boldmath Measurement of ${\cal{B}}(\upsln\to\chi_{cJ}+X$)}
	
	We search for $\upstochicx$ by reconstructing the 
$\chi_{cJ}\to\jpsi\gamma$ decay. Photon candidates are
required to have energy $E_{\gamma}>$100 MeV, not be matched to a charged 
track and have a shower shape consistent with that of a photon.
We also require that the invariant mass of this photon with any other
photon in the event is greater than 2.5 standard deviations away from the 
$\pi^0$ mass of 135 MeV/$c^2$~\cite{pdg}.
Photon candidates passing these selection criteria are combined with
$\jpsi$ candidates to form a $\chi_c$ candidate. As done previously, we 
compute mass differences, $M(l^+l^-\gamma)-M(l^+l^-)$, 
for $\jpsi$ candidates which have a mass in the 
range $3.00<M(l^+l^-)<3.14$ GeV/$c^2$. From this distribution, we subtract the analogous
distribution obtained from the $\jpsi$ sidebands, here defined as candidates with
$2.88<M(l^+l^-)<2.95$ GeV/$c^2$ or $3.20<M(l^+l^-)<3.27$ GeV/$c^2$. 
As was done for the $\psitwos$, we  combine and average the 
$\jpsimm$ and $\jpsiee$ channels. The
invariant mass difference distribution, $M_{l^+l^-\gamma}-M_{l^+l^-}$, 
is shown in Fig.~\ref{chic_fit_fixedwid}.
The solid histogram is the $\upsln$ data and the shaded histogram is
the $\upsl4$ continuum, scaled by the ratio of luminosities. The $\upsln$ data
is fit using three Gaussians on top of an exponential background. The Gaussian
means are restricted to lie within 5 MeV/$c^2$ of the world average 
values of the differences between the $\chi_{cJ}$ and $\jpsi$ 
masses (see Table~\ref{chic_to_jpsigamma})~\cite{pdg} and the widths 
are constrained to the values found from simulation:
12.5 MeV/$c^2$, 10.5 MeV/$c^2$ and 10.3 MeV/$c^2$ for $J=0,1,2$.
The larger width for $J=0$ is a result of the 16.2 MeV intrinsic width which
is included in the simulation.  
The fitted yields are 0$\pm$13, 52$\pm$12, and 47$\pm$11 events for
$J$ = 0, 1, 2, respectively. The significance of each signal is given 
by $S/\sqrt{B}$, where $S$ is the signal yield, and $B$ is
the estimated background within $\pm$3 standard deviations of the
fitted mean using the exponential background function. The significances
are found to be 3.9 and 4.1 for the $J$=1 and $J$=2 states, respectively.
The averaged efficiencies for
the $\jpsimm$ and $\jpsiee$ channels are (27$\pm$1)\%, (30$\pm$1)\%
and (28$\pm$1)\% for $J$ = 0, 1, 2 states, respectively. The efficiency
for reconstructing the $J=0$ state includes a $\sim$2\% loss of signal 
events due to events in the tails of the Breit-Wigner.
Branching fractions 
for $\upstochicx$ are computed relative to $\upstojpsix$ and are tabulated 
in Table~\ref{chic_to_jpsigamma}. The last column 
shows the measured fraction of $\jpsi$'s which come from $\chi_{cJ}$ feed-down,
which is (11$\pm$3)\% for the $J=1$ state and (10$\pm$2)\% for the $J=2$ state.
We only obtain upper limits on the $J=0$ state.  Theoretical estimates
of this ratio are at the level of 10\% for the sum of all three
$\chi_{cJ}$ states~\cite{cheung_1s,trottier}. The rates we report here are 
higher than those expectations.

\begin{figure}[btp]
\centerline{
\includegraphics[width=3.5in]{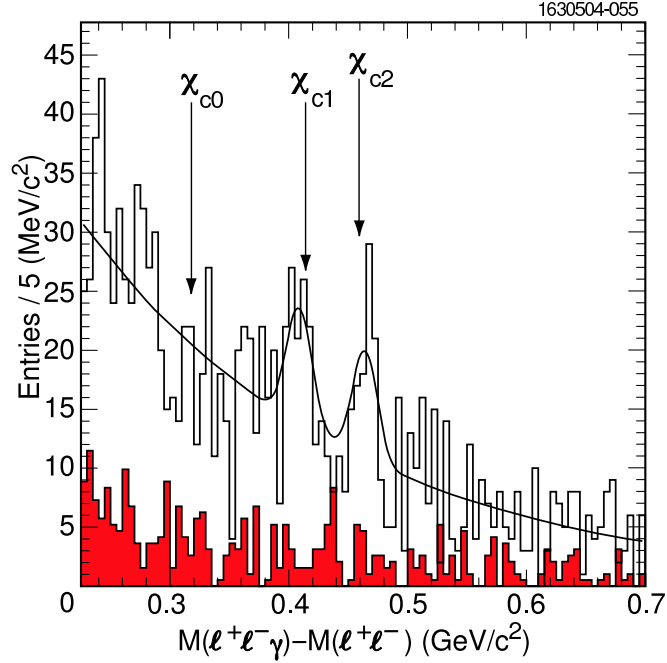}}
\caption{\label{chic_fit_fixedwid}
Difference of invariant masses, $M(l^+l^-\gamma)-M(l^+l^-)$, for
$M(l^+l^-)$ in the range from 3.00 to 3.14 GeV/$c^2$
for $l=\mu,e$ combined.
The solid histogram represents the $\upsln$ data and the shaded histogram
is the below-$\upsl4$ data scaled by the ratio of luminosities. The arrows indicate
the mass differences corresponding to the three $\chi_c$ states.} 
\end{figure}

\begin{table*}[hbt] 
\begin{center}  
\caption{Measurements of the branching fractions for $\upstochicx$ with the
relevant inputs. The table includes, from left to right, the $\chi_{cJ}$ states,
the world average mass difference $M_{\chi_{cJ}}-M_{\jpsi}$, the branching
fractions ${\cal{B}}(\chi_{cJ}\to\jpsi\gamma)$, the observed yields, the
reconstruction efficiencies, the computed branching fractions relative to
$\upstojpsix$, and the computed
feed-down to $\jpsi$. For the $\chi_{c0}$ we show 90\% confidence level upper 
limits\label{chic_to_jpsigamma}.} 
\begin{tabular}{|c|c|c|c|c|c|c|}\hline 
Mode & $M_{\chi_{cJ}}-M_{\jpsi}$ & ${\cal{B}}(\chi_{cJ}\to\jpsi\gamma$) & $N_{\rm events}$ & Eff.(\%) 
& ${{\cal{B}}(\upsln\to\chi_{cJ}+X)\over {\cal{B}}(\upstojpsix)}$ & Feed-down\\ 
   & (MeV/$c^2$)  & (\%) & ($\mu\mu$+$ee$) &  ($\mu\mu$+$ee$)  & ($\mu\mu$+$ee$)  & to $\jpsi$ \\ 
\hline 
$\chi_{c0}$ &   318  &1.11$\pm$0.15   &  0$\pm$13 & (27$\pm$1)\%  & $< 5.9$ & $<$0.065 \\
$\chi_{c1}$ &   414  &31.6$\pm$2.7   & 52$\pm$12 & (30$\pm$1)\%  & 0.35$\pm$0.08 & 0.11$\pm$0.03 \\
$\chi_{c2}$ &   459  &20.2$\pm$2.0   & 47$\pm$11 & (28$\pm$1)\%  & 0.52$\pm$0.12 & 0.10$\pm$0.02 \\
\hline
\end{tabular} 
\end{center} 
\end{table*}

\subsection{\boldmath Cross-check using $B\to\chi_{cJ}+X$ in $\upsl4$ Data}

	As a consistency check, we measure the branching fraction for
${\cal{B}}(B\to\chi_{cJ}+X)$ using 10.4 million $B$ decays from the $\upsl4$ data
sample. We restrict the $\chi_{cJ}$ to have momentum less than 1.6 GeV/$c$, which is 
the kinematic limit from $B$-meson decay. A clear signal is found only 
for the $J=1$ state, for which the fitted yield is 347$\pm$35 events. The
continuum background is negligible, and is therefore neglected.
The reconstruction efficiency is determined using a $B\to\chi_{cJ}+X$ MC
simulation, and is found to be (28$\pm$1)\%. The branching fraction for
${\cal{B}}(B\to\chi_{c1}+X)$ is thus found to be
$(3.3\pm 0.3({\rm stat}))\times 10^{-3}$, which is consistent with the 
PDG value of $(3.6\pm0.3)\times10^{-3}$~\cite{pdg}.
The 90\% confidence level upper limit on ${\cal{B}}(B\to\chi_{c2}+X)$ is 
1.5$\times10^{-3}$, which does not conflict with the measured branching fraction of 
(1.3$\pm$0.4)$\times10^{-3}$~\cite{pdg,chic2}. 

\section{\boldmath Estimates of Systematic Uncertainty}
\label{sec:sys}

\subsection{\boldmath Uncertainties in $\upstojpsix$}

	The branching fractions for $B\to\jpsi+X$ using $\jpsimm$ and
$\jpsiee$ were shown in Section~III.I to be higher than
the world average values by 7\% for $\jpsimm$ and 5\% for $\jpsiee$,
which is taken as the systematic uncertainty in reconstructing
these decays. We ascribe an additional uncertainty due to our limited
knowledge of the final state in $\upstojpsix$ and its modeling.
%in the CLEO Monte Carlo program, {\sc qq}~\cite{qq}.
This additional uncertainty is taken to be
half the difference in the reconstruction efficiency obtained from
our $\upstojpsix$ simulation and 
that obtained using the $B\to\jpsi+X$ simulation. This results 
in additional contributions
of 4\% for $\jpsimm$ and 6\% for $\jpsiee$. We also
include an additional 5\% uncertainty in each due to limited MC statistics.
We therefore estimate 9\% systematic uncertainty in the $\jpsimm$ 
reconstruction efficiency and 8\% for $\jpsiee$.

	The uncertainty in the signal yield is estimated by floating the
Gaussian widths used in fitting each $x$ bin. We find that the
signal yield changes by 3\% for both the $\jpsimm$ and $\jpsiee$ analyses.
The systematic error due to uncertainty in the shape of the background was estimated
by comparing a linear background shape with an exponential. The difference
is found to be 2\% for $\jpsimm$ and 1\% for $\jpsiee$.

	Systematic uncertainty in the background subtraction comes from
lack of precise knowledge of the continuum cross section ($\eetojpsix$) and
the error in the extrapolation from the $\upsl4$ energy to the
$\upsln$ energy. The latter includes uncertainties in the ratio
of luminosities and the branching fraction ${\cal{B}}(\upstomm)$.
Our measurement of the rate for $\eetojpsix$ is
uncertain at the level of 10\% (statistical uncertainty only) and 
the extrapolation factor of 1.41$\pm$0.18
is uncertain at the level of 13\%. We therefore estimate that the overall
background rate is uncertain at the level of 16\%. This uncertainty is
propagated to an error in the branching fraction by
shifting the central value for the
background (see Fig.~\ref{y4s_x_dist}) up and then down by one standard 
deviation, and in each case, computing the change in the branching fraction
from the nominal value. The corresponding shift
in the branching fraction for $\upstojpsix$ is found to be 6\% for both
$\jpsimm$ and $\jpsiee$, which is taken 
as the systematic uncertainty due to the background subtraction. 
The uncertainty on the background subtraction does not have a significant 
effect on the general shape of the momentum distribution since its 
contribution is only about 10\% of the total.

	We also include the uncertainty in the number of $\upsln$ decays, 
which is estimated to be 1\% based on the number of $\upsln$ events and
the uncertainty in the off-to-on $\upsln$ luminosity ratio. We also include 
a 2\% relative uncertainty in ${\cal{B}}(\jpsi\to l^+l^-)$. The total uncertainty 
is therefore found to be 12\% for the $\jpsimm$ channel and 11\% for $\jpsiee$
channel. 

The systematics are itemized and shown in columns 2 and 3 in Table~\ref{sys}.

\subsection{\boldmath Uncertainties in $\upsln\to\psitwos+X$ Analysis}

	For the $\psitwos$ (as well as the $\chi_{cJ}$) analysis, 
many of the systematic
uncertainties cancel since these measurements are 
reported as a ratio with respect to the $\upstojpsix$ branching fraction. 
The uncertainty in the $\psitwos$ reconstruction efficiency comes
from limited MC statistics (5\%) and an imperfect understanding 
of $\upsln\to\psitwos+X$ events. The uncertainty from the latter
is taken to be the half the difference between the efficiency
obtained using our default $\upsln\to\psitwos+X$ simulation and
the $B\to\psitwos+X$ simulation. The two simulations agree to within 1\% in
absolute value,
which translates into an additional 5\% relative systematic uncertainty in
the reconstruction efficiency for each channel. We also include a
systematic uncertainty of 1\% per track for each of the two pions
in the decay $\psitwos\to\jpsi\pi^+\pi^-$ (2\%). We therefore
estimate that the non-cancelling systematic uncertainty in the
$\psitwos$ reconstruction efficiency is 7\%.

	Uncertainty in the signal yield is estimated by 
shifting the Gaussian width up and down by 20\% about the central 
value (2.5 MeV/$c^2$) and taking half of the average deviation,
which results in a 7\% systematic uncertainty.
Uncertainty in signal yield due to the assumed background shape was estimated
by fitting the background to the alternate functional form: 
$A(1-B\exp(-Cx))$ (the default is a second-order polynomial).
The yield differs by 6\%, which is taken as the associated 
uncertainty.
	
	Uncertainty due to the background subtraction
is estimated by considering a $\pm$50\% change in the expected
background contribution, which is about two standard deviations. 
The resulting systematic uncertainty in the branching ratio is 15\%.
Uncertainty in the branching fraction 
${\cal{B}}(\psitwos\to\psi\pi^+\pi^-)$ contributes 5\%. We therefore estimate 
a total systematic uncertainty of 20\% in the $\psitwos$ branching fraction
ratio.

\subsection{\boldmath Uncertainties in $\upstochicx$ Analyses}

	The uncertainty in the efficiencies for reconstructing
$\chi_{cJ}\to\jpsi\gamma$ is taken to be half the difference in the
efficiencies for reconstructing $\chi_{cJ}$ in $B$ decays at the 
$\upsl4$ versus in $\upsln$ decays (8\%). We include an additional
uncertainty of 8\% to reflect the lower value we obtain for 
${\cal{B}}(B\to\chi_{c1}+X)$ as compared to the world average.
This also accounts for any possible systematic uncertainty in the photon 
reconstruction efficiency.
We attribute a 3\% uncertainty for each due to limited MC statistics.
We therefore estimate an overall systematic uncertainty in the reconstruction efficiency
of the $\chi_{cJ}+X$ final state of 12\%. 

	The uncertainty in the signal
yield is obtained by allowing the Gaussian widths to float, from which we obtain
differences of 6\% and 1\% for the $J=1$ and $J=2$ states. The uncertainty
from the background determination is estimated by using different ranges
over which to fit the background. We find maximum variations of 12\%, of
which we take half as the associated systematic uncertainty (6\%). Since there
is no evidence of any $\chi_{cJ}$ signal in the continuum, we do not
consider this as a source of systematic uncertainty. 
Lastly, we include uncertainties in the $\chi_{cJ}\to\jpsi\gamma$ branching 
fractions~\cite{pdg} of 20\%, 10\% and 11\% for $J$ = 0, 1 and 2 states, respectively.
We therefore estimate total systematic uncertainties of 18\% for $\chi_{c1}$ 
and 17\% for $\chi_{c2}$. The uncertainty on the limits for $\chi_{c0}$ are 
estimated to be 25\%, and the upper limits are increased by this amount to 
reflect this systematic uncertainty.

	Systematics uncertainties are listed and summarized in Table~\ref{sys}.

\begin{table*}[hbt] 
\begin{center}  
\caption{Sources of systematic uncertainty in the $\upstojpsix$, 
$\upsln\to\psitwos+{\rm{X}}$ and $\upsln\to\chi_{cJ}+{\rm{X}}$ analyses.\label{sys}}
\begin{tabular}{|c|c|c|c|c|c|c|}
\hline 
Source 			& \multicolumn{6}{|c|}{Value (\%)} \\
\hline
          	        & $\jpsimm$ & $\jpsiee$ & $\psitwos$ & $\chi_{c0}$ & $\chi_{c1}$ & $\chi_{c2}$ \\
\hline
Reconstruction efficiency 	& 9\%   & 8\%    & 7\%	& 12\%	& 12\%	& 12\%	\\
Signal Yield		  	& 3\%   & 3\%   & 7\% 	&  -    & 6  & 1    \\ 
Background Shape  		& 2\%  	& 1\%   & 6\%	&   -   & 6  & 6 \\ 
Background subtraction  	& 6\%  	& 6\%   & 15\%	&   -   & -  & - \\ 
\#$\upsln$ decays 		& 1\%   & 1\%   & -     & -     & -  & - \\
Error in ${\cal{B}}(\jpsi\to l^+ l^-)$ & 2\% & 2\% & - & - & -   & - \\
Error in ${\cal{B}}(\psitwos\to\psi\pi^+\pi^-)$ & - & - & 5\% & - & - & - \\
Error in ${\cal{B}}(\chi_{c0,1,2}\to\jpsi\gamma)$ & - & - & - & 20\% & 10\% & 11\% \\
\hline 
Total & 12 & 11 & 20 & 25 & 18 & 17\\
\hline
\end{tabular} 
\end{center} 
\end{table*} 

\section{\boldmath Summary}
\label{sec:summary}

	We present vastly improved measurements of the rates for production of
charmonium in $\upsln$ decays over previous measurements. We have measured 
both the branching fraction for $\upstojpsix$ and the scaled momentum 
distribution, as well as distributions in
the polar angle and helicity in $\upsln$ decay.
We also report on first observations of the decays $\upsln\to\psitwos+X$ and
evidence for $\upsln\to\chi_{c1,2}+X$.

	The branching fractions for $\upstojpsix$ are
measured in both the $\jpsimm$ and $\jpsiee$ channels. Their braching 
fractions, ${\cal{B}}(\upstojpsix)$ are measured to be 
$(6.9\pm0.5({\rm stat})\pm0.8({\rm syst}))\times 10^{-4}$ and
$(6.1\pm0.5({\rm stat})\pm0.7({\rm syst}))\times 10^{-4}$, respectively.
The two are averaged to obtain:

\begin{widetext}
\begin{equation}
{\cal{B}}(\upstojpsix) = (6.4\pm0.4({\rm stat})\pm0.6({\rm syst}))\times 10^{-4} \nonumber \\
\end{equation}
\end{widetext}

	We also measure the branching fraction  
${\cal{B}}(\upsln\to\psitwos+X)$ relative to
${\cal{B}}(\upstojpsix)$, and find

\begin{widetext}
\begin{equation}
{{\cal{B}}(\upsln\to\psitwos+X)
\over {\cal{B}}(\upsln\to\jpsi+X)} = 0.41\pm0.11({\rm stat})\pm0.08({\rm syst}). 
\nonumber \\
\end{equation}
\end{widetext}

	This report also presents the first evidence of the decay 
$\upsln\to\chi_{cJ}+X$. The branching fractions for all measured modes are 
summarized in Table~\ref{table:summary}. 

\begin{table*}[hbt] 
\begin{center}  
\caption{Summary of measurements of the branching fractions for $\upsln$
to charmonium final states. The $\psitwos$ and $\chi_{cJ}$ branching fractions are
expressed relative to ${\cal{B}}(\upstojpsix)$.
The last column show the feed-down contributions
to $\upstojpsix$. For the $\chi_{c0}$ we show upper limits at 90\% confidence
level. Where uncertainties are shown, the first is statistical and the second
is systematic\label{table:summary}.} 

\begin{tabular}{|c|c|c|}\hline 
Final state, $f$ & ${\cal{B}}(\upsln\to f+X)$ & Feed-down to $\jpsi$ \\
\hline
$\jpsi$     & $(6.4\pm0.4\pm0.6)\times 10^{-4}$ & -  \\
\hline
            & ${\cal{B}}(\upsln\to f+X)$/${\cal{B}}(\upstojpsix)$ & \\
\hline
$\psitwos$  & $0.41\pm0.11\pm0.08$ & $0.24\pm0.06\pm0.05$ \\
$\chi_{c0}$ & $<$7.4 & $<$0.082 \\
$\chi_{c1}$  & $0.35\pm0.08\pm0.06$ & $0.11\pm0.03\pm0.02$ \\ 
$\chi_{c2}$  & $0.52\pm0.12\pm0.09$ & $0.10\pm0.02\pm0.02$ \\ 
\hline
\end{tabular} 
\end{center} 
\end{table*} 

	The $\upstojpsix$ branching fraction is consistent with
predictions of both the color-octet mechanism for $\jpsi$ production in 
$\upsln$ decays~\cite{cheung_1s,napsuciale} and
color-singlet production via $\upstojpsiccg$~\cite{li-xie-wang}, each 
which predict a branching fraction at the level of 6$\times 10^{-4}$.
The observed scaled momentum spectrum is relatively soft, peaking 
around $x\simeq0.3$, in contrast to $\jpsi$'s produced in the continuum, which 
peak at about 0.7. The peaking at low momentum is in sharp contrast to the
prediction of the color-octet model which predicts a peaking of $x$ near 1.
It is possible that incorporation of final state interactions could improve 
this agreement as
was shown for $\eetojpsix$~\cite{fleming1}. The observed spectrum is 
closer to, although softer than, the expectation of the color-singlet 
process~\cite{li-xie-wang}, $\upstojpsiccg$, which peaks near $x\simeq$0.5.
When this parton-level calculation is simulated using {\sc pythia} we
are able to achieve satisfactory agreement in the region $x<0.6$ when
hadronization of the recoiling charm quarks into charm hadrons is
included, and our measured feed-downs of $\psitwos$ and $\chi_{cJ}$ 
to $\jpsi$ are incorporated.

	The observation of $\upsln\to\psitwos+X$ is
the first to a $c\bar{c}$ final state other than $\jpsi$. The feed-down to
$\jpsi$ constitutes (24$\pm$6$\pm$5)\% of the inclusive rate for $\upstojpsix$, 
which is significantly 
larger than expected in either the color-octet~\cite{cheung_1s} or
color-singlet model~\cite{li-xie-wang}, each which predict a feed-down 
to $\jpsi$ at the level of 10\%.
Our measured rates for $\upstochicx$ yield feed-down contributions of  
(11$\pm$3$\pm$2)\% for the $J$=1 state and (10$\pm$2$\pm$2)\% 
for the $J$=2 state, which is also
larger than the expected contribution of about 10\%, summed
over $J$=0, 1, and 2~\cite{cheung_1s,trottier,li-xie-wang}.

	These measurements can shed additional light on the
role of the color-octet and color-singlet mechanisms 
in producing charmonium, not only in
$\upsln$ decays but also in $e^+e^-$ and $p\bar{p}$ collisions.
In this regard, it would be of great interest to 
determine whether the same softening mechanism applied to the color-octet
prediction for $\eetojpsix$~\cite{fleming1} can account for the 
$\jpsi$ momentum spectrum in $\upstojpsix$. Moreover, computation of the
angular distributions for the color-octet and color-singlet mechanisms
may provide additional discrimination between these two processes. 
From an experimental perspective, the additional information on the
roles of color-singlet versus 
color-octet mechanisms may be obtained by measuring the ratio 
$\sigma(p\bar{p}\to\jpsi c\bar{c}+X)/\sigma(p\bar{p}\to\jpsi+X)$
at the Tevatron. The unexpectedly large value for 
$\sigma(e^+e^-\to\jpsi c\bar{c}+X)/\sigma(e^+e^-\to\jpsi+X)$ reported by
Belle~\cite{belle_ee_jpsi2} may point to a large rate in $p\bar{p}$
collisions as well.

	We gratefully acknowledge the effort of the CESR staff in providing 
us with excellent luminosity and running conditions.
This work was supported by the National Science Foundation, and the
U.S. Department of Energy. We also thank Kingman Cheung and Wai-Yee Keung
for their assistance with the color-octet predictions and Shi-yuan Li
for providing color-singlet predictions.

\end{document}